\documentclass[a4paper,amsmath,amssymb,11pt,eqsecnum]{revtex4}
\usepackage{mathtools}
\usepackage{graphicx}
\usepackage[breaklinks, colorlinks, citecolor=blue]{hyperref}
\usepackage{amsmath}
\usepackage{amssymb}
\usepackage{bm}% bold math
\usepackage{verbatim}
\usepackage{subfigure}

\graphicspath{{./}}

\def\half{\frac{1}{2}}
\newcommand{\partdif}[2]{\frac{\partial {#1}}{\partial {#2}}}
\newcommand{\funcdif}[2]{\frac{\delta {#1}}{\delta {#2}}}
\newcommand{\figref}[1]{Fig.~\ref{#1}}

\newcommand{\sgn}[1]{\operatorname{sgn} (#1)}

\begin{document}

\begin{flushleft}
    KCL-PH-TH/2014-34
\end{flushleft}

\title{Fourth order deformed general relativity}

\author{Rhiannon Cuttell} \email{rhiannon.cuttell@kcl.ac.uk}
\author{Mairi Sakellariadou} \email{mairi.sakellariadou@kcl.ac.uk}
\affiliation{
	Department of Physics,
	King's College London,
	University of London,
	Strand,
	London,
	WC2R 2LS,
	U.K.
}

\begin{abstract}
Whenever the condition of anomaly freedom is imposed within the framework of effective approaches to loop quantum cosmology, one seems to conclude that a deformation of general covariance is required. Here, starting from a general deformation we regain an effective gravitational Lagrangian including terms up to fourth order in extrinsic curvature.  We subsequently constrain the form of the corrections for the homogeneous case, and then investigate the conditions for the occurrence of a big bounce and the realisation of an inflationary era, in the presence of a perfect fluid or scalar field.
\end{abstract}

\maketitle

\tableofcontents

%%%%%%%%%%%%%%%%%%%%%%%%%%%%%%%%%%%%%%%%%%%%%%%%%%%%%%%%%%

\section{Introduction}
\label{sec:intro}

There are many equivalent ways of formulating classical general relativity.  Although covariant methods are arguably the most pure, since they explicitly retain the general space-time covariance
(coordinate invariance), difficulties can arise when trying to apply them to certain physical systems.  For example, trying to combine covariant general relativity with quantum mechanics is problematic, since the nature of time is very different for both theories.

To bring general relativity more in line with the way quantum mechanics is usually formulated, one can use canonical methods which split the space-time structure to a spatial hypersurface that evolves over time \cite{bojowald2010canonical}. Canonical general relativity can be formulated equivalently using different variables.  There is geometrodynamics, which uses the spatial metric and extrinsic curvature $(q_{ab}, K^{ab})$; connection dynamics, which uses the Ashtekar connection and densitised triads $(A^i_a, E^a_i)$; and loop dynamics, which uses holonomies of the connection and gravitational flux $(h_{\gamma}[A],F^i_{\gamma}[E])$.  Classically, $h_{\gamma}[A]$ is given by the path-ordered exponential of the connection integrated along a curve $\gamma$ and $F^i_{\gamma}[E]$ is the flux of the densitised triad through a surface that the curve $\gamma$ intersects. If we take $\gamma$ to be infinitesimal we can easily relate loop dynamics and connection dynamics because then $h_{\gamma} = 1 + A(\dot{\gamma}) + \mathcal{O}(|\gamma|^2)$ \cite{Rovelli2014}.

However, loop quantum gravity pictures space-time as not being a continuous manifold, but being composed of a network of nodes connected by ordered links with quantum numbers for geometrical quantities such as volume.  Such a network is not embedded in space but \emph{is space itself}.  As such, one cannot shrink the length of a link between nodes to being infinitesimal as in the classical case, and so the relationship between loop dynamics and connection dynamics is broken due to the quantisation of geometry.

If general relativity is truly the classical limit of loop quantum gravity, then there should be a semi-classical limit where the dynamics are well approximated by general relativity with small effective quantum corrections.  At small scales and high curvature, these corrections should become important.

When general relativity is formulated using canonical methods, how is the space-time general covariance retained when there is an explicit splitting of space and time?  The spatial general covariance of the hypersurface coordinates and the invariance under different embeddings of the hypersurface in space-time are given by different constraint equations.  The former is given by the diffeomorphism constraint $D^a$, and the latter by the Hamiltonian constraint $H$, and both must weakly vanish for physical solutions.  Depending on the choice of canonical variables, there may also be a Gauss constraint $G^i$, though it is not related to the space-time structure.  The choice of coordinates \emph{should} be equivalent to a gauge choice as long as the constraints vanish, but satisfying the constraint equations is not quite enough.  The constraints also need to satisfy certain interrelations.

The classical constraints form a closed Poisson bracket algebra \cite{bojowald2010canonical}.  The constraints are the generators of deformations of the hypersurface (equivalent to coordinate transformations or evolution through time) and therefore if the algebra were not closed due to anomalies $A_{IJ}$, namely,
\begin{equation}
    \{C_I, C_J\} =
    f^{K}_{IJ} C_K + A_{IJ},
    \qquad
    C_I \in \{H,D^a,G^i\},
    \qquad
    A_{IJ} \notin \{H,D^a,G^i\},
\end{equation}
one could show that a spatial hypersurface which satisfies the constraints at one time will not satisfy the constraints at all times.  Therefore the anomalies $A_{IJ}$ must strongly vanish because the algebra is required to be closed for the theory to be consistent.

The interpretation of general relativity as a geometric theory of space-time and whether our spatial hypersurface can be embedded in space-time is intimately related to the specific form of this algebra, as shown in Ref.~\cite{teitelboim_how_1973}.  This form, written for the smeared versions of the constraints, is
\begin{subequations}
\begin{align}
    \{ D[N_{1}^{a}], D[N_{2}^{b}] \} & =
    D[ \mathcal{L}_{N_{2}} N_{1}^a ],
        \label{eq:conalg_dd}\\
    \{ H[N_{1}], D[N_{2}^{a}] \} & = 
    H[ \mathcal{L}_{N_{2}} N_{1} ],
        \label{eq:conalg_hd}\\
    \{ H[N_1], H[N_2] \} & =
    D[ q^{ab} (N_1 \partial_b N_2 - N_2 \partial_b N_1) ],
    \label{eq:conalg_hh}
\end{align}
    \label{eq:conalg}%
\end{subequations}
where $N$ and $N^a$ are the lapse and shift functions, respectively, which are gauge quantities that specify the embedding of the spatial hypersurface in space-time \cite{bojowald2010canonical}. The $q^{ab}$ denotes the spatial metric, while $H[N]$ is the smearing of $H$ with $N$ over the hypersurface, and $\mathcal{L}_X Y$ is the Lie derivative of $Y$ with respect to the vector $X^a$.

When quantising general relativity, these constraints are promoted to operators, and they satisfy a commutator algebra corresponding to Eq.~\eqref{eq:conalg}.  Classical constraints satisfy $C_{I} \approx 0$ and their quantum versions satisfy $\hat{C_{I}} |\Psi \rangle = 0$. Note that the form of the algebra must not contain anomalies when the constraints are not satisfied (i.e. when `off-shell'), since off-shell states can become important in quantum mechanics (e.g., virtual intermediate states may influence particle scattering). 

Loop quantum cosmology attempts to be a symmetry-reduced form of loop quantum gravity \cite{bojowald_loop_2004, Ashtekar2006, sakellariadou_lattice_2007-1}.  The derivation of it from the full theory has not yet been done, so it proceeds by quantising mini-superspace models using methods gained from developments in the full theory.  Effective approaches to loop quantum cosmology work in a semi-classical scheme, and try to include quantum corrections to the classical theory which model two main features \cite{bojowald_quantum_2012}.  One comes from the underlying primary nature of holonomies that cannot be made infinitesimal and so must be approximated through higher-order curvature terms and non-localities. The other is due to the inverse-volume operator, which is present in the Hamiltonian constraint and will produce corrections at small scales.  This is because the volume operator includes zero in its spectrum and the inverse-volume operator cannot have eigenvalues which are infinite, so the effective approach applies a cut-off function to regularise the zero-volume limit. 

Investigations into the consistency of these effective models of loop quantum cosmology have shown that imposing the algebra to be non-anomalous produces a modification to the algebraic structure functions.  In particular, Eq.~\eqref{eq:conalg_hh} must be modified by a phase space functional $\beta [q_{ab}, K^{ab}]$, determined by quantum corrections, (see Refs.~\cite{Cailleteau2012a, Cailleteau2013} and references in Ref.~\cite{bojowald_deformed_2012}), leading to
\begin{equation}
    \{ H[N_1], H[N_2] \} = D[ \beta q^{ab} (N_1 \partial_b N_2 - N_2 \partial_b N_1) ].
    \label{eq:modconalg_hh}
\end{equation} 
Hence, in such case the effective Hamiltonian constraint $H$ is modified, but the diffeomorphism constraint $D^a$ is not.  The interpretation of this is that the structure within the spatial hypersurface is the same as in general relativity, but the structure involving the time-like direction (i.e. the embedding of the hypersurface) is not.

Since the algebra is modified but without anomalies, the symmetry underlying our models must not be space-time general covariance but deformed to a related and more broad kind of symmetry.  There are relations between this ``deformed general relativity'' and the so-called {\sl deformed special relativity} \cite{magueijo_lorentz_2002, magueijo_generalized_2003}, a phenomenological model seeking quantum gravitational deformations to the Poincar\'e symmetry group such that the Planck scale becomes observer independent.  In some versions of the deformed special relativity, the dispersion relation for particles is deformed, and a particle's speed becomes dependent on its energy (i.e. an energy-dependent speed of light).  In effective loop quantum cosmology, the quantum corrections can alter the speed of propagation of electromagnetic and gravitational waves \cite{bojowald_loop_2008}, but a dispersion relation for individual particles, similar to the one obtained within deformed special relativity, has not been found so far.

Note that there are very strong observational constraints on a variable speed of light and there exist theoretical problems with its implications of locality becoming a relative concept \cite{hossenfelder_box-problem_2009}. It has been argued that only a variable speed of light dependent on local energy density or curvature would be consistent with observer independence, but this means that for individual particles a difference in time-of-flight would be unobservable \cite{hossenfelder_multi-particle_2007}. 

However, even if these corrections do not produce observable effects for individual particles, they may have important implications for cosmology.  A prediction of loop quantum cosmology is that the big bang singularity, unavoidable in classical gravity, is resolved being replaced by a big bounce \cite{bojowald_loop_2004, Ashtekar2006}.  In Ref.~\cite{Cailleteau2012a} the form of the correction function $\beta$ is obtained for scalar perturbations around an isotropic and homogeneous background, whilst including holonomy corrections $\beta = \cos{2K}$ (modulo a few constants), where $K$ is the extrinsic curvature.  This implies that for situations of high curvature such as during the very early universe, the sign of $\beta$ can change.  The algebra of constraints in Eq.~\eqref{eq:conalg} is only true for Lorentzian manifolds, and the sign of Eq.~\eqref{eq:conalg_hh} is reversed for Euclidean manifolds.  Whilst space-time in this regime could certainly not be interpreted in terms of a classical Euclidean manifold, this ``signature change'' implies that propagation of information ceases since the time-like direction becomes space-like. Note that this has also been called ``asymptotic silence'' and seems to occur also in other approaches to quantum gravity \cite{mielczarek_asymptotic_2012,ambjorn_quantum_2013}.  To some extent, this might be a concrete mechanism of realising something like the Hartle-Hawking no-boundary proposal \cite{Hartle1983}. 

One may regain a constraint algebra that has its general relativistic form (and therefore be coordinate invariant) by making a canonical change of variables.  In Ref.~\cite{tibrewala_inhomogeneities_2013}, a specific case was considered and it was shown that the variables that satisfied this had absorbed the quantum corrections into their definition. However, this may not be possible to do for a system which is not already symmetry-reduced.

In Section~\ref{sec:regainlag}, we find an effective gravitational Lagrangian for loop quantum cosmology by starting from the modified constraint algebra.  We then see how our results relate to previous investigations in loop quantum cosmology in Section~\ref{sec:cosmo} and calculate the conditions for either a bounce or inflation to occur. We summarise our results in Section~\ref{sec:concl}, while some technical parts are presented in an appendix.

%%%%%%%%%%%%%%%%%%%%%%%%%%%%%%%%%%%%%%%%%%%%%%%%%%%%%%%%%%

\section{Regaining an effective Lagrangian}
\label{sec:regainlag}

As Hojman, Kucha\u{r} and Teitelboim showed in Ref.~\cite{hojman_geometrodynamics_1976}, for general relativity just as it is possible to derive the form of the constraint algebra \eqref{eq:conalg} by specifying the Hamiltonian and diffeomorphism constraints, it is possible to derive the form of the Hamiltonian constraint by specifying the form of the constraint algebra and the diffeomorphism constraint.  Kucha\u{r} also showed in Ref.~\cite{kuchar_geometrodynamics_1974} how to derive the form of the gravitational Lagrangian from the same starting point.  This was extended by Bojowald and Paily in Ref.~\cite{bojowald_deformed_2012}, where they started from the deformed algebra \eqref{eq:modconalg_hh} and derived the most general effective Lagrangian that satisfies it up to second order in extrinsic curvature.  In this section, we extend this further to include up to fourth order terms in extrinsic curvature. In Appendix~\ref{sec:nonlocal} we perform the first derivation of an effective Lagrangian when including a specific version of corrections (spatial holonomies) which would imply non-local effects if no expansion were performed.  In a derivative expansion, these corrections (while becoming local) modify the classical expression.  Performing a truncated local expansion up to second order in derivatives we find that no non-local effects appear. In Section~\ref{sec:cosmo}, we show that including up to fourth order terms in extrinsic curvature leads to the appearance of a big bounce, as often found in loop quantum cosmology. 

Let us emphasise that in the present study we only consider spatial derivatives appearing up to linear order in the spatial Ricci curvature $R = ^{(3)}\!\!R$, and we leave for a future investigation the case where higher-order spatial derivatives, higher-order time derivatives and non-linearities in spatial derivatives are included \cite{Deruelle2010,bojowald_discreteness_2014}.  Let us also emphasise that after regaining an effective Lagrangian, we will only analyse the background equations. 

%%%%%%%%%%%%%%%%%%%%%%%%%%%%%%%%%%%%%%%%%%%%%%%%%%%%%%%%%%%%%%%%%%

\subsection{2nd order}
\label{subsec:regainlag_2nd}

Instead of the extrinsic curvature, the independent variable we will use as the ``velocity'' is $v_{ab} := N^{-1} \{ q_{ab}, H[N] \}$, which is the flow of the metric normal to the spatial foliation. Classically, this is equal to twice the extrinsic curvature, 
\begin{equation}
	v_{ab} = 2K_{ab} = \frac{1}{N} \left( \dot{q}_{ab} - 2 N_{(a|b)} \right),
\end{equation} 
and since it depends on our choice of coordinates through $N$ and $N^a$, it is fairly arbitrary (we can choose  $N^a$ so that $v_{ab} = 0$, i.e. a static coordinate system).  If there are deformations of space-time structure (i.e. $\beta \neq 1$), the quantity $\half v_{ab}$ may no longer be able to be interpreted as geometrical extrinsic curvature \cite{bojowald_deformed_2012}.

Let us begin by outlining the way to get a second order effective Lagrangian.  We can use Eq.~\eqref{eq:modconalg_hh} to find 
\begin{equation}
	\funcdif{L(x)}{q_{ab}(y)} v_{ab}(y) + \beta(x) D^{a}(x) \delta_{|a}(x,y)
	- (x \leftrightarrow y) = 0,
\label{eq:modconalg_hh_deriv}
\end{equation}
where $X_{|a}$ denotes the covariant derivative of $X$ which is compatible with the spatial metric, $q_{ab|c} = 0$.  The spatial structure should not be modified, and therefore neither should the diffeomorphism constraint (since it generates spatial transformations), so we substitute the usual formula $D^a = -2p^{ab}_{|b}$ into the second term of Eq.~\eqref{eq:modconalg_hh_deriv} (where $p^{ab}$ is the canonical momentum of the metric).  After multiplying by test functions, integrating by parts, discarding total derivatives, and substituting $p^{ab} := \partial L / \partial v_{ab}$, we find the useful distribution equation,
\begin{equation}
	\funcdif{L(x)}{q_{ab}(y)} v_{ab}(y) + 2 \beta_{|b}(x) \partdif{L(x)}{v_{ab}(x)} \delta_{|a}(x,y) + 2 \beta(x) \partdif{L(x)}{v_{ab}(x)} \delta_{|ab}(x,y) - (x \leftrightarrow y) = 0.
	\label{eq:dist_eqn}
\end{equation}
We can regain an effective Lagrangian by expanding the Lagrangian and the correction function in powers of $v_{ab}$, namely
\begin{subequations}
\begin{align}
	L(x) & = \sum_{n=0}^{\infty} L^{i_1 j_1 \ldots i_n j_n}
             [q_{ab}] v_{i_1 j_1}(x) \ldots v_{i_n j_n} (x),
		\label{eq:local_lagrangian_expansion} \\
	\beta(x) & = \sum_{n=0}^{\infty} \beta^{i_1 j_1 \ldots
                  i_n j_n} [q_{ab}] v_{i_1 j_1}(x) \ldots v_{i_n j_n}
                (x).
		\label{eq:local_correction_expansion}
\end{align}
	\label{eq:local_expansion}%
\end{subequations}
This expansion is valid for local corrections such as those from inverse-triad quantisation and from local effects of holonomy quantisation \cite{bojowald_deformed_2012}.  Holonomy corrections arise from the fact that the holonomy phase-space variables in loop quantum gravity result from integrating connections along a path, and
are therefore in general non-local in character.  To properly include non-local effects in our expansion, we need to expand in terms of spatial derivatives of $v_{ab}$, which is what we do in Appendix~\ref{sec:nonlocal} below.  In general, we should also take into account non-localities in time, which would involve higher time
derivatives being taken into account (such as the metric acceleration), but we consider that is beyond the scope of this paper.

Equation~(\ref{eq:dist_eqn}) was used by Bojowald and Paily in Ref.~\cite{bojowald_deformed_2012} to derive the effective Lagrangian to second order,
\begin{equation}
	L = \frac{\sqrt{\det q}}{2 \kappa} \left(
		\frac{{\rm sgn}(\beta^{\varnothing})}{\sqrt{|\beta^{\varnothing}|}}
		\frac{v^{ab} v_{ab} - (v^a_a)^2}{4}
		+ \sqrt{|\beta^{\varnothing}|} R - 2 \Lambda
	\right),
	\label{eq:lagrangian_2nd}
\end{equation}
which, in the classical limit $\beta^{\varnothing} = 1$, becomes the standard Arnowitt-Deser-Misner Lagrangian. Only the $v$-independent part of the correction function $\beta$, denoted by $\beta^{\varnothing}$, appears at second order of the effective action.  Note that $R$ is the spatial Ricci scalar, $\Lambda$ is the cosmological constant and $\kappa = 8 \pi G$.  

In deriving the above effective Lagrangian to second order, we obtain the following relations, which we will use later:
\begin{subequations}
\begin{gather}
	L^{ab}_{|b}=0,
		\label{eq:lagrangian_identities_1} \\
	2 L^{abcd}_{|b} \beta^{\varnothing} + L^{abcd} \beta^{\varnothing}_{|b} = 0.
		\label{eq:lagrangian_identities_2}
\end{gather}
	    \label{eq:lagrangian_identities}%
\end{subequations}
The former is obtained from the fact that generically $\beta^{\varnothing}\neq 0$ and the latter is deduced from the fact that  $L^{\varnothing}(x)$, as a spatial scalar density, is to second order in spatial derivatives the function 
$L^{\varnothing}(x)=L^{\varnothing}(q_{ij}(x),^{(3)}\!R_{ij}(x))$.

%%%%%%%%%%%%%%%%%%%%%%%%%%%%%%%%%%%%%%%%%%%%%%%%%%%%%%%%%%%%

\subsection{3rd order}
\label{subsec:regainlag_3rd}

To better facilitate calculations involving tensors with many indices, we are going to adopt a convention where we can write a pair of symmetric indices $a_1 a_2$ as $A$, so that $L^A := L^{(a_1 a_2)} = L^{a_1 a_2}$.  For example, Eq.~\eqref{eq:lagrangian_identities_2} can be re-written as $2 L^{AB}_{|a_2} \beta^{\varnothing} + L^{AB}
\beta^{\varnothing}_{|a_2}$=0.  We do this because the coefficients in the expansion \eqref{eq:local_expansion} are symmetric only under permutation of \emph{pairs} of their indices (also, each pair of indices is itself symmetric).

We substitute the local expansion \eqref{eq:local_expansion} into Eq.~\eqref{eq:dist_eqn} and collect terms which are quadratic in $v_{ab}$ and its spatial derivatives,
\begin{equation}
\begin{split}
	0  & = 
	\funcdif{L^{A}(x)}{q_{B}(y)} v_{A}(x) v_{B} (y)
        \\
	& + 2\left[
		L^{A} ( \beta^{BC} v_{B} v_{C} )_{|a_1}
		+ 2 L^{AB} v_{B} ( \beta^{C} v_{C} )_{|a_1}
		+ 3 L^{ABC} v_{B} v_{C} \beta^{\varnothing}_{|a_1}
	\right]^{(x)} \delta_{|a_2} (x,y)
        \\
	& +  2\left[
		L^{A} \beta^{BC}
		+ 2 L^{AB} \beta^{C}
		+ 3 L^{ABC} \beta^{\varnothing}
	\right]^{(x)} v_{B}(x) v_{C}(x) \delta_{|A} (x,y)
	- (x \leftrightarrow y),
\end{split}
	\label{eq:dist_eqn_3rd}
\end{equation}
where the superscript $(x)$ means that all terms within the brackets are functions of $x$ only.

%%%%%%%%%%%%%%%%%%%%%%%%%%%%%%%%%%%%%%%%%%%%%%%%%%%%%%

\subsubsection{Test functions}
\label{subsubsec:regainlag_3rd_test_function}

Following the method used in
Refs.~\cite{kuchar_geometrodynamics_1974,bojowald_deformed_2012}, we multiply \eqref{eq:dist_eqn_3rd} by test functions $a(x)$ and $b(y)$, then integrate by parts over $x$ and $y$, note which terms disappear due to symmetry of indices, discard total derivatives, and use \eqref{eq:lagrangian_identities_1} to get
\begin{equation}
\begin{split}
	0 & = \int {\rm d}^3 x \, {\rm d}^3 y \left\{
			\funcdif{L^{A}(x)}{q_{B}(y)} - \funcdif{L^{B}(y)}{q_{A}(x)}
		\right\} a(x) b(y) v_{A}(x) v_{B}(y)
        \\
	& - 2\int {\rm d}^3 x (a b_{|a_1} - a_{|a_1} b)^{(x)} \left[
			2 (L^{AB} v_{B})_{|a_2} \beta^{C} v_{C}
			+ 3 (L^{ABC} v_{B} v_{C})_{|a_2} \beta^{\varnothing}
		\right]^{(x)}.
\end{split}
	\label{eq:test_function_3rd}
\end{equation}
Setting $a = b = 1$ (therefore $a_{|a} = b_{|a} = 0$), we obtain
\begin{equation}
	\left(
		\funcdif{L^{A}(x)}{q_{B}(y)} - \funcdif{L^{B}(y)}{q_{A}(x)}
	\right) v_{A}(x) v_{B}(y) = 0.
\end{equation}
Following Ref.~\cite{kuchar_geometrodynamics_1974}, we therefore deduce that
\begin{equation}
	\funcdif{L^{A}(x)}{q_{B}(y)} - \funcdif{L^{B}(y)}{q_{A}(x)} = 0,
\end{equation}
meaning that the ``functional curl'' of $L^{A}(x)$ vanishes.  This
condition, combined with \eqref{eq:lagrangian_identities_1}, leads to
\begin{equation}
    L^{ab} = \sqrt{\det q} \left[ \theta_q q^{ab} + \theta_R \left(
  R^{ab} - \half q^{ab} R \right) \right],
	\label{eq:lagrangian_form_1}
\end{equation}
where $\theta_q$ and $\theta_R$ are constants.

Using Eq.~\eqref{eq:test_function_3rd} and setting $v_{ab|c} = 0$, we obtain that $2 L^{AB}_{|a_2} \beta^{C} + 3
L^{ABC}_{|a_2} \beta^{\varnothing}$ must vanish independently:
\begin{subequations}
	\begin{equation}
		2 L^{AB}_{|a_2} \beta^{C} + 3 L^{ABC}_{|a_2} \beta^{\varnothing} = 0.
		\label{eq:lagrangian_identities_3a}
	\end{equation}
The remaining part which needs to vanish is
    \begin{equation}
		L^{AB} \beta^{C}
		+ 3 L^{ABC} \beta^{\varnothing} = 0.
		\label{eq:lagrangian_identities_3b}
	\end{equation} 
Combining Eqs.~(\ref{eq:lagrangian_identities_3a}),
(\ref{eq:lagrangian_identities_3b}) and using
\eqref{eq:lagrangian_identities_2}, leads to 
    \begin{equation}
        \left( L^{ABC} \beta^{\varnothing} \right)_{|a_2} = 0,
		\label{eq:lagrangian_identities_3c}
	\end{equation}
which, from \eqref{eq:lagrangian_identities_3b}, implies 
	\begin{equation}
		(L^{AB} \beta^{C})_{|a_2} = 0.
		\label{eq:lagrangian_identities_3d}
	\end{equation}
	    \label{eq:lagrangian_identities_3}%
\end{subequations}
%
%%%%%%%%%%%%%%%%%%%%%%%%%%%%%%%%%%%%%%%%%%%%%%%%%%%%%%

\subsubsection{Consistency check}
\label{subsubsec:regainlag_3rd_consistency}

We will show that the terms in the correction function and Lagrangian expansions which do not satisfy time-reversal invariance (i.e. are proportional to an odd power of extrinsic curvature) are all
required to disappear for consistency requirements.  

Consider the symmetries of Eq.~\eqref{eq:lagrangian_identities_3b}:
\begin{equation}
	L^{(AB)} \beta^{C} + 3 L^{(ABC)} \beta^{\varnothing} = 0,
	\label{eq:lagrangian_identities_3b_sym}
\end{equation}
and then symmetrise Eq.~\eqref{eq:lagrangian_identities_3b_sym} by adding all permutations of the indices,
\begin{equation}
	\frac{1}{3} \left(L^{(AB)} \beta^{C} + L^{(BC)} \beta^{A} + L^{(CA)} \beta^{B} \right)
	 + 3 L^{(ABC)} \beta^{\varnothing} =0,
\end{equation}
which is, equivalently written,
\begin{equation}
    L^{(AB} \beta^{C)} + 3 L^{(ABC)} \beta^{\varnothing} = 0.
\end{equation}
Then, by substituting  Eq.~\eqref{eq:lagrangian_identities_3b_sym}, 
one finds
\begin{equation}
	L^{(AB)} \beta^{C} = L^{(AB} \beta^{C)}.
\end{equation}
Let us write this out explicitly,
\begin{equation}
    3 L^{(AB)} \beta^{C} = L^{(AB)} \beta^{C} + L^{(BC)} \beta^{A} + L^{(CA)} \beta^{B},
\end{equation}
cancel the first term on the right hand side, and then contract this with $q_{a_1 b_1} q_{a_2 b_2}$, to obtain
\begin{equation}
	2 L^{ab}_{\;\;\;ab} \beta^{cd} = 2 L^{cd}_{\;\;\;ab} \beta^{ab}.
    \label{L^ab}
\end{equation}
Equation~\eqref{eq:lagrangian_2nd} then implies $L^{abcd} \propto q^{a(c} q^{d)b} - q^{ab} q^{cd}$, which combining with Eq.~\eqref{L^ab} leads to
\begin{equation}
	2 \beta^{ab} = - q^{ab} \beta^c_c.
    \label{beta^ab}
\end{equation}
Contracting Eq.~\eqref{beta^ab} with $q_{ab}$, one finds that $2 \beta^a_a = -3 \beta^a_a$, which clearly means that $\beta^a_a = 0$. Hence
\begin{equation}
	\beta^{ab} = 0.
	\label{eq:beta_1st_condition}
\end{equation} 
This means that no deformation of the constraint algebra which violates time-reversibility at the linear level can be consistent. 
If we compare this to Eq.~\eqref{eq:lagrangian_identities_3b}, we see that $L^{ABC} = 0$ since $\beta^{\varnothing}$ is generically non-zero; and if we compare it to the result calculated in the appendix, \eqref{eq:lagrangian_identities_3extra}, we see that
\begin{equation}
	L^{A} \beta^{BC} = 0,
	    \label{eq:lag1orbeta2vanish}
\end{equation} 
and so either the first order Lagrangian term or the second order correction term must vanish.  Since we set out to look for higher-order terms and would prefer for time-reversal symmetry to be respected, we take this to imply that $L^{A} = 0$.  From Eq.~\eqref{eq:lagrangian_form_1}, this means that $\theta_{q} = \theta_{R} = 0$.

%%%%%%%%%%%%%%%%%%%%%%%%%%%%%%%%%%%%%%%%%%%%%%%%%%%%%%%%%%%%%%%%%%%%%%%%%%

\subsection{4th order}
\label{subsec:regainlag_4th}

Let us return to Eq.~\eqref{eq:dist_eqn} and collect terms cubic in $v_{ab}$ and its spatial derivatives. We thus get
\begin{equation}
\begin{split}
    0 & = \funcdif{L^{AB}(x)}{q_{C}(y)} v_{A}(x) v_{B}(x) v_{C}(y)
    + 2\delta_{|a_1}(x,y) \bigg[ L^{A} (\beta^{BCD} v_{B} v_{C} v_{D})_{|a_2}
        \\ &
        + 2 L^{AB} v_{B} ( \beta^{CD} v_{C} v_{D} )_{|a_2}
        + 3 L^{ABC} v_{B} v_{C} (\beta^{D} v_{D})_{|a_2}
  		+ 4 L^{ABCD} v_{B} v_{C} v_{D} \beta^{\varnothing}_{|a_2} \bigg]^{(x)}
  		\\ &
  	+ 2\delta_{|A}(x,y) \bigg[
		L^{A} \beta^{BCD}
		+ 2 L^{AB} \beta^{CD}
		+ 3 L^{ABC} \beta^{D}
		+ 4 L^{ABCD} \beta^{\varnothing}
	\bigg]^{(x)} v_{B}(x) v_{C}(x) v_{D}(x)
	    \\ &
	- (x \leftrightarrow y).
\end{split}
	\label{eq:dist_eqn_4th}
\end{equation}
Let us first multiply by test functions $a(x)$ and $b(y)$ and then integrate by parts over $x$ and $y$, note which terms disappear due to symmetry of indices, discard total derivatives, and use  Eq.~\eqref{eq:lagrangian_identities_1} to finally get 
\begin{equation}
\begin{split}
	0 & = \int {\rm d}^3 x \,
      {\rm d}^3 y \, a(x) \, b(y) \left\{
      \funcdif{L^{AB}(x)}{q_{C}(y)} v_{A}(x) v_{B}(x) v_{C}(y)
      -\funcdif{L^{AB}(y)}{q_{C}(x)} v_{A}(y) v_{B}(y) v_{C}(x)
      \right\}
        \\
	& - 2 \int {\rm d}^3 x (a \, b_{|a_1} - a_{|a_1} b)^{(x)} \bigg[
		2 (L^{AB} v_{B})_{|a_2} \beta^{CD} v_{C} v_{D}
		+ 3 (L^{ABC} v_{B} v_{C})_{|a_2} \beta^{D} v_{D}
	    \\
	& \hspace{4em}  
	    + 4 (L^{ABCD} v_{B} v_{C} v_{D})_{|a_2} \beta^{\varnothing}	\bigg]^{(x)},
\end{split}
	\label{eq:test_function_4th} 
\end{equation}
and for the same reasons as for Eq.~\eqref{eq:test_function_3rd}, the first and second integrals must vanish independently.  Focusing on the second integral and setting $v_{ab|c} = 0$ we are left with
\begin{subequations}
	\begin{equation} 2L^{AB}_{|a_2} \beta^{CD} + 3 L^{ABC}_{|a_2} \beta^{D} +
        4 L^{ABCD}_{|a_2} \beta^{\varnothing} = 0,
		\label{eq:lagrangian_identities_4a}
	\end{equation}
and the other part which must also vanish is
	\begin{equation} L^{AB} \beta^{CD} + 3 L^{ABC} \beta^{D} + 6 L^{ABCD}
        \beta^{\varnothing} = 0.
		\label{eq:lagrangian_identities_4b}
	\end{equation}
Combining the last two equations and using Eq.~\eqref{eq:lagrangian_identities_2}, we get
	\begin{equation} 2 L^{ABCD}_{|a_2} \beta^{\varnothing} + 3 L^{ABCD}
        \beta^{\varnothing}_{|a_2} = 0.
		\label{eq:lagrangian_identities_4c}
	\end{equation}
Equation~\eqref{eq:beta_1st_condition} implies that $\beta^D$ vanishes, hence Eq.~\eqref{eq:lagrangian_identities_4b} reads
	\begin{equation} L^{AB} \beta^{CD} + 6 L^{ABCD} \beta^{\varnothing} = 0,
	    \label{eq:lagrangian_identities_4d}
	\end{equation}
    	\label{eq:lagrangian_identities_4}%
\end{subequations}
and since we know the form of the second order term, we can just rearrange this equation to write down the fourth order term:
\begin{equation}
	L^{ABCD} = \frac{-1}{6 \beta^{\varnothing}}L^{AB} \beta^{CD}.
\end{equation}
Hence, to fourth order, our effective Lagrangian is
\begin{equation}
	L = \frac{\sqrt{\det q}}{2 \kappa} \left[
		\frac{{\rm sgn}(\beta^{\varnothing})}{\sqrt{|\beta^{\varnothing}|}}
		\frac{v^{ab} v_{ab} - (v^a_a)^2}{4} \left(
			1 - \frac{1}{6 \beta^{\varnothing}} \beta^{cdef} v_{cd} v_{ef}
		\right)
		+ \sqrt{|\beta^{\varnothing}|} R - 2 \Lambda
	\right].
	\label{eq:lagrangian_4th}
\end{equation}
Notice that it appears that, at the level of the action, we cannot absorb the correction through a simple redefinition of the form $v'_{ab} = f(\beta) v_{ab}$.

%%%%%%%%%%%%%%%%%%%%%%%%%%%%%%%%%%%%%%%%%%%%%%%%%%%%%%%%%%%%%%%%%%%%

\subsubsection{Consistency check}
\label{subsubsec:regainlag_4th_further}

To take this even further, and constrain the form that $\beta^{abcd}$ can take, let us go back to \eqref{eq:lagrangian_identities_4}, and look at the symmetries like we did previously in sub-section~\ref{subsubsec:regainlag_3rd_consistency}. By symmetrising the whole equation, we have
\begin{equation}
	L^{(AB)} \beta^{(CD)} = L^{(AB} \beta^{CD)}.
\end{equation}
If we then expand this, and do some simple algebraic manipulations, we get
\begin{equation}
	5 L^{(AB)} \beta^{(CD)} - L^{(AC)} \beta^{(BD)} - L^{(AD)} \beta^{(BC)}
	- L^{(BC)} \beta^{(AD)} - L^{(BD)} \beta^{(AC)} - L^{(CD)} \beta^{(AB)} = 0,
\end{equation}
which one we contract with $q_{a_1 b_1} q_{a_2 b_2} q_{c_1 d_1} q_{c_2 d_2}$, and ignore an overall multiplication
factor, leads to
\begin{equation}
	L^{ab}_{\;\;\;ab}\, \beta^{cd}_{\;\;\;cd} - L^{abcd} \beta_{abcd} = 0.
\end{equation} 
Then since $L^{abcd} \propto q^{a(c} q^{d)b} - q^{ab} q^{cd}$, the above equation implies
\begin{equation} 2 \beta^{ab}_{\;\;\;ab} + \beta^{a \;\;
  b}_{\;\;a \;\; b} = 0.
	\label{eq:correction_identities_2b}
\end{equation}
Equation~\eqref{eq:lagrangian_identities_4a}, keeping in mind that the first order correction term vanishes, implies
\begin{equation}
    L^{AB}_{|a_2} \beta^{CD} + 2 L^{ABCD}_{|a_2} \beta^{\varnothing} = 0,
\end{equation}
which combined with Eq.~\eqref{eq:lagrangian_identities_4d} leads to
\begin{equation}
    -6 (L^{ABCD} \beta^{\varnothing})_{|a_2} - L^{AB} \beta^{CD}_{|a_2} + 2L^{ABCD}_{|a_2} \beta^{\varnothing} = 0.
\end{equation}
If we now expand the first term, and then use the identity \eqref{eq:lagrangian_identities_4c}, we find
\begin{equation}
    L^{AB} \beta^{CD}_{|a_2} = 0,
\end{equation}
which, since $L^{AB} \neq 0$, leads us to the conclusion that
\begin{equation}
    \beta^{abcd}_{|e} = 0.
	\label{eq:correction_identities_2c}
\end{equation}

Each of the correction expansion coefficients in \eqref{eq:local_correction_expansion} is a function of the metric
only.  Out of all forms we can find the metric in ($\det q$, $q_{ab}$, $q_{ab|c}$, \dots), the only ones which can be included in a tensor which has a vanishing uncontracted covariant derivative is the metric itself and its determinant.  If we combine this information with \eqref{eq:correction_identities_2b}, we find that
\begin{equation}
	\beta^{abcd} [q_{ij}] = \beta^{(2)}[\det q] (q^{a(c}q^{d)b} - q^{ab} q^{cd}).
\end{equation}
The metric determinant is a scalar density with a non-zero weight, but $\beta^{(2)}$ must simply be a scalar.  Classically this would mean that $\beta^{(2)}$ would have to be a constant, but since we are dealing with a semi-classical effective theory we expect that there will be quantum degrees of freedom in the full theory which may be able to balance the weight. Note that the same argument holds for $\beta^{\varnothing}$ \cite{bojowald_deformed_2012}.

Hence, the effective Lagrangian to fourth order reads
\begin{equation}
	L = \frac{\sqrt{\det q}}{2 \kappa} \left[
		\frac{{\rm sgn}(\beta^{\varnothing})}{\sqrt{|\beta^{\varnothing}|}}
		\frac{v^{ab} v_{ab} - (v^a_a)^2}{4}
		-\frac{\beta^{(2)}}{|\beta^{\varnothing}|^{3/2}}
		\frac{\left(v^{ab} v_{ab} - (v^a_a)^2\right)^2}{24}
		+ \sqrt{|\beta^{\varnothing}|} R - 2 \Lambda
	\right].
	\label{eq:lagrangian_4th_further}
\end{equation}
From this, we can find an equation for the metric momentum through the canonical formula $p^{ab} = \funcdif{L}{v_{ab}}$.  This however gives a relation which is cubic in $v_{ab}$, and so it may be possible to invert it only locally to get an equation for the Hamiltonian, $H = v_{ab} p^{ab} - L$ which only depends on $q_{ab}$ and $p^{ab}$, for certain ranges of $v_{ab}$ (or equivalently for certain ranges of extrinsic curvature).  However, the obtained relationship may be complicated to apply usual canonical methods to our results, and one must use variational methods.

Since general covariance is not explicit because of separating space and time derivatives, this makes calculations very complicated for systems which we have not already symmetry-reduced. However, it may be possible to perform a similar calculation to find an effective form for $H$ that is fourth order in $p^{ab}$ by following the procedure in Ref.~\cite{hojman_geometrodynamics_1976}, which would then allow canonical methods to be used.

%%%%%%%%%%%%%%%%%%%%%%%%%%%%%%%%%%%%%%%%%%%%%%%%%%%%%%%%%%%%%%%%%%%%%%%%%%%%

\subsection{Discussion}
\label{subsec:regainlag_discussion}

We have constructed an effective Lagrangian which includes the first correction terms of higher extrinsic curvature from a generally deformed constraint algebra, which may stem from holonomy effects of loop dynamics.  The only term in the modified algebra of constraints which was deformed is the $\{H,H\}$ term which, since the Hamiltonian constraint generates time-like translations, modifies the time structure of space-time.  Teitelboim in
Ref.~\cite{teitelboim_how_1973} showed how the algebra of constraints in general relativity is related to the ability to embed a space-like hypersurface into a space-time with geometric interpretation.  This algebra is deformed, hence the interpretation of space-time in terms of classical geometry breaks down.  Since we are only using an effective geometrodynamic approximation to the underlying quantum geometry, break down of the classical geometry should be expected. 
%In our scheme, the increased dependence on extrinsic curvature suggests a physical dependence on our definition of time - something not in keeping with the spirit of general relativity or loop quantum gravity.

One issue with higher-order theories of classical gravity such as certain $F\left(^{(4)}\!R\right)$ theories, is that they can often suffer from ghosts and are thus unstable.  In our case, since \eqref{eq:lagrangian_4th_further} contains non-linearities in the time derivatives of the form $\dot{q}^4$, one may fear that this may also be the case here.  However, since we are only dealing with an effective model, such a situation is likely to simply be a relic of our truncation of the curvature expansion.

%%%%%%%%%%%%%%%%%%%%%%%%%%%%%%%%%%%%%%%%%%%%%%%%%%%%%%%%%%%%%%%%%%%%

\section{Cosmology}
\label{sec:cosmo}

Our investigation is primarily directed towards finding possible phenomenological effects of loop dynamics.  We expect observable corrections to physical dynamics to only be present in extreme systems, such as during the era of high energy density in the early universe.  Thus, in this section we investigate the cosmological implications of our effective scheme.  This has of course been studied for loop quantum cosmology in the past, but our effective scheme may allow for greater flexibility when studying the phenomenology.

\subsection{Background equations}
\label{subsec:cosmo_background}

We restrict to a flat Friedmann-Lema\^{i}tre-Robertson-Walker (FLRW) space with $\Lambda = 0$,
\begin{equation}
    \bar{L} = \frac{3 a^3}{\kappa \bar{N}^2 \sqrt{\bar{\beta}^{\varnothing}}} \mathcal{H}^2 \left( 1 + \frac{4 \bar{\beta}^{(2)}}{\bar{N}^2 \bar{\beta}^{\varnothing}} \mathcal{H}^2 \right),
	\label{eq:lagrangian_4th_cosmo}
\end{equation}
where $a$ is the scale factor, $\mathcal{H} = \dot{a}/a$ is the Hubble parameter, and an overbarred function means just the background component of that function (i.e. only dependent on $a$).

We couple this to matter with energy density $\rho$ and pressure density $P = w \rho$.  We Legendre transform the effective Lagrangian to find the Hamiltonian.  Imposing the Hamiltonian constraint $H \approx 0$ gives us 
\begin{equation}
    \mathcal{H}^2 \left( 1 + \frac{12 \bar{\beta}^{(2)}}{\bar{N}^2 \bar{\beta}^{\varnothing}} \mathcal{H}^2 \right) = \frac{\kappa \bar{N}^2 \sqrt{\bar{\beta}^{\varnothing}}}{3} \rho,
\end{equation}
which can be solved to find the modified Friedmann equation,
\begin{equation}
    \mathcal{H}^2 = \frac{2 \kappa \bar{N}^2 \sqrt{\bar{\beta}^{\varnothing}}}{3(x+1)} \rho,
	\label{eq:mod_friedmann}
\end{equation}
where the correction factor is
\begin{equation} x := \sqrt{1+\frac{16 \kappa
    \bar{\beta}^{(2)}}{\sqrt{\bar{\beta}^{\varnothing}}} \rho}.
	\label{eq:correction_factor}
\end{equation}
Going back to the effective Lagrangian, and varying it with respect to the scale factor, we find the Euler-Lagrange equation of motion.  When we substitute in Eq.~\eqref{eq:mod_friedmann}, we get the acceleration
equation
\begin{equation}
\begin{split}
    \frac{\ddot{a}}{a} = \frac{-\kappa \bar{N}^2 \sqrt{\bar{\beta}^{\varnothing}}}{6x} \rho & \bigg\{ 1+3w - 2 \partdif{\,\ln \bar{N}}{\,\ln a} - \half \partdif{\,\ln \bar{\beta}^{\varnothing}}{\,\ln a}
        \\
    & -2 \left( \frac{x-1}{x+1} \right) \left[ 1+ \partdif{\,\ln \bar{N}}{\,\ln a} - \half \partdif{}{\,\ln a}\ln \left( \frac{\bar{\beta}^{(2)}}{\bar{\beta}^{\varnothing}} \right) \right] \bigg\}.
	\label{eq:mod_acceleration}
\end{split}
\end{equation}
If we take the time derivative of Eq.~\eqref{eq:mod_friedmann}, then substitute in Eq.~\eqref{eq:mod_acceleration}, we get the usual continuity equation
\begin{equation}
	\dot{\rho} + 3 \mathcal{H} \rho (1+w) = 0.
	\label{eq:continuity}
\end{equation}
Note that there may be corrections to the matter sector due to the modified constraint algebra \cite{bojowald_radiation_2007, bojowald_quantum_2013}, but we have not included these here.

Since $\beta^{(2)}$ vanishes in the classical limit, we can treat it as a small parameter to expand Eq.~\eqref{eq:mod_friedmann} to first order, 
\begin{equation}
    \mathcal{H}^2 = \frac{\kappa \bar{N}^2 \sqrt{\bar{\beta}^{\varnothing}}}{3} \rho \left( 1 - \frac{\rho}{\rho_c} \right) + \mathcal{O} \left( \frac{\rho^2}{\rho_c^2} \right),
	\label{eq:mod_friedmann_bounce}
\end{equation}
where
\begin{equation}
    \rho_c := \frac{\sqrt{\bar{\beta}^{\varnothing}}}{4 \kappa \bar{\beta}^{(2)}},
	\label{eq:critical_density}
\end{equation}
and expanding the bracket in Eq.~\eqref{eq:mod_acceleration} to first order, we find that $\ddot{a}/a>0$ when $w<w_a$, where
\begin{equation}
    w_a = \frac{-1}{3} \left[ 1 - \half \partdif{\,\ln \bar{\beta}^{\varnothing}}{\,\ln a} -2 \frac{\rho}{\rho_c} \left[ 1 - \half \partdif{}{\,\ln a}\ln \left( \frac{\bar{\beta}^{(2)}}{\bar{\beta}^{\varnothing}} \right) \right] \right].
	    \label{eq:eqn_state_acceleration}
\end{equation}
We have set $\bar{N}=1$, so this is applicable for cosmic time.

The modified Friedmann equation \eqref{eq:mod_friedmann_bounce} predicts a big bounce rather than a big bang, since $\dot{a} \to 0$ as $\rho \to \rho_c$.  This requires either $\rho_c$ to be constant, or for it to diverge at a slower rate than $\rho$ as $a \to 0$.

Let us emphasise that the bounce is found considering only holonomy corrections manifesting as higher-order powers of curvature and ignoring higher-order terms in the derivative expansion. The equations \eqref{eq:mod_friedmann_bounce} and \eqref{eq:eqn_state_acceleration} have been expanded to leading order in $\bar{\beta}^{(2)}$, so we should be cautious about the regime of their validity.  Noting that the Lagrangian is also an expansion; $\bar{\beta}^{(2)}$ is a coefficient to the fourth order term and appears only linearly, we conclude that there is no good reason why we should have more trust in equations such as \eqref{eq:mod_friedmann} or \eqref{eq:mod_acceleration} simply because they contain higher orders.  In Ref.~\cite{Ashtekar2006}, Ashtekar,
Pawlowski and Singh write their effective Friedmann equation with leading order corrections (which is the same as
\eqref{eq:mod_friedmann_bounce}) and say that it holds surprisingly well even for $\rho \approx \rho_c$, the regime when the expansion should break down (we should note that their work refers only to the case where $w = 1$).

%%%%%%%%%%%%%%%%%%%%%%%%%%%%%%%%%%%%%%%%%%%%%%%%%%%%%%%%%%%%%%%%%%%%%

\subsection{\texorpdfstring{$\bar{\beta}$}{Background deformation} functions}
\label{subsec:cosmo_beta}

We need to know $\bar{\beta}^{\varnothing} (a)$ and $\bar{\beta}^{(2)} (a)$ in order to make progress beyond this point, so we compare our results to those found in previous investigations.  In Ref.~\cite{Cailleteau2013}, Cailleteau, Linsefors and Barrau have found information about the correction function for when inverse-volume and holonomy effects are both included in a perturbed FLRW system.  Their equation (Eq.~$(5.18)$ in Ref.~\cite{Cailleteau2013}) gives (rewritten slightly)
\begin{equation}
    \bar{\beta}(a,\dot{a}) = f(a) \Sigma(a,\dot{a}) \partdif{^2}{\dot{a}^2} \left[ \gamma^{\varnothing}(a,\dot{a}) \left( \frac{\sin[\gamma \mu(a) \dot{a}]}{\gamma \mu(a)} \right)^2 \right],
        \label{eq:beta_barrau_general}
\end{equation}
where $\gamma \approx 0.12$ is the Barbero-Immirzi parameter, $\gamma^{\varnothing}$ is the function which contains information about inverse-volume corrections, $\Sigma(a,\dot{a})$ depends on the form of $\gamma^{\varnothing}$, and $f(a)$ is left unspecified.  We just consider the case where $\gamma^{\varnothing}=\gamma^{\varnothing}(a)$, in which case $\Sigma = 1/\left(2\sqrt{\gamma^{\varnothing}}\right)$ and $\mu = a^{2\omega}\sqrt{\gamma^{\varnothing} \Delta}$ with $\omega = -1/2$. The constant $\Delta$ is usually interpreted as being the ``area gap'' derived in loop quantum gravity.  We leave $\omega$ unspecified for now, because different quantisations of loop quantum cosmology give it
equal to different values in the range $[-1/2,0]$.
Equation~\eqref{eq:beta_barrau_general} now becomes
\begin{equation}
	\bar{\beta} = f \sqrt{\gamma^{\varnothing}}
		\cos \left(
			2 \gamma \sqrt{\gamma^{\varnothing} \Delta} a^{\delta} \mathcal{H}
		\right),
	\label{eq:beta_barrau_specific}
\end{equation}
where $\delta = 1+2\omega$.  The ``old dynamics'' or ``$\mu_0$ scheme'' corresponds to $\omega = 0$ and $\delta = 1$, and the favoured ``improved dynamics'' or ``$\bar{\mu}$ scheme'' corresponds to $\omega = -1/2$ and $\delta = 0$ \cite{sakellariadou_lattice_2007-1,   sakellariadou_lattice_2007}.  In the semi-classical regime, $\sqrt{\Delta} \mathcal{H} \ll 1$, so we can Taylor expand this equation for the correction function to get
\begin{equation}
	\bar{\beta} \approx f \sqrt{\gamma^{\varnothing}}
	 - 2 \gamma^2 \Delta a^{2\delta} f (\gamma^{\varnothing})^{3/2} \mathcal{H}^2.
	 \label{eq:beta_barrau_expand}
\end{equation}
The way that $\gamma^{\varnothing}$ is defined is that it multiplies the background gravitational term in the Hamiltonian constraint relative to the classical form.  Since we are assuming $\gamma^{\varnothing} = \gamma^{\varnothing}(a)$, we can isolate it by taking our Lagrangian \eqref{eq:lagrangian_4th_cosmo} and setting
$\bar{\beta}^{(2)}=0$.  If we then Legendre transform to find a Hamiltonian expressed in terms of the momentum of the scale factor, we find that it is proportional to $\sqrt{\bar{\beta}^{\varnothing}}$. Thus, we conclude that $\bar{\beta}^{\varnothing} = \left( \gamma^{\varnothing} \right)^2$ when $\gamma^{\varnothing}$ is just a function of the scale factor.  Using this to compare \eqref{eq:beta_barrau_expand} with what we have already found for our correction function,
\begin{equation}
	\bar{\beta} \approx \bar{\beta}^{\varnothing}
		+ \bar{\beta}^{(2)} \left[ \bar{v}_{ab} \bar{v}^{ab}
		- \left( \bar{v}^a_a \right)^2 \right]	
	= \bar{\beta}^{\varnothing} - 24 \bar{\beta}^{(2)} \mathcal{H}^2,
	    \label{eq:beta_background}
\end{equation}
we find that $f = \left(\bar{\beta}^{\varnothing}\right)^{3/4}$, and therefore $f = (\gamma^{\varnothing})^{3/2}$.
From this, we can now deduce the form of the coefficient for the higher-order corrections,
\begin{equation}
	\bar{\beta}^{(2)} = \frac{\gamma^2 \Delta}{12} a^{2\delta} ( \gamma^{\varnothing} )^3.
\end{equation}
The exact form of $\gamma^{\varnothing}(a)$ is uncertain, and the possible forms that have been found also contain quantisation ambiguities.  The form given by Bojowald in Ref.~\cite{bojowald_loop_2004} is
\begin{equation}
	\gamma^{\varnothing} = \frac{3r^{1-l}}{2l} \left\{
		\frac{(r+1)^{l+2}-|r-1|^{l+2}}{l+2} -r
		\frac{(r+1)^{l+1}-\sgn{r-1}|r-1|^{l+1}}{l+1} \right\},
\end{equation} 
where $l\in (0,1)$, $r = a^2 / a_{\star}^2$ and $a_{\star}$ is the characteristic scale of the inverse-volume corrections, related to the discreteness scale.  We will only use the asymptotic expansions of this function, namely 
\begin{equation}
	\gamma^{\varnothing} \approx \left\{
\begin{aligned}
	& 1 + \frac{(2-l)(1-l)}{10} \left( \frac{a}{a_{\star}} \right)^{-4},
    & {\rm if } \; a \gg a_{\star}
        \\
    & \frac{3}{1+l} \left( \frac{a}{a_{\star}} \right)^{2(2-l)},
    & {\rm if } \; a
        \ll a_{\star}
\end{aligned}
    \right.
	    \label{eq:inv_vol_cases}
\end{equation}
and even then we will only take $\gamma^{\varnothing} \approx 1$ for $a \gg a_{\star}$, since the correction is vanishingly small.

In Planck units, $\hbar = c = 1$ and $l_{\rm Pl}^2 = m_{\rm Pl}^{-2} = G$.  We replace the area gap with a dimensionless parameter $\tilde{\Delta} = \Delta l_{\rm Pl}^{-2}$ which is of order unity.  Our modified Friedmann equation is now given by
\begin{equation}
	\mathcal{H}^2 =
	\frac{8 \pi \gamma^{\varnothing}}{3 m_{\rm Pl}^2} \rho \left[
		1 - \frac{8 \pi \gamma^2 \tilde{\Delta}}{3} a^{2\delta}
		(\gamma^{\varnothing})^2 \frac{\rho}{\rho_{\rm Pl}} \right].
	\label{eq:mod_friedmann_planck}
\end{equation}
We will apply this to different types of matter.  First of all we will consider a perfect fluid, and then we will consider a scalar field with a power-law potential.

%%%%%%%%%%%%%%%%%%%%%%%%%%%%%%%%%%%%%%%%%%%%%%%%%%%%%%%%%%%%

\subsection{Perfect fluid}
\label{subsec:cosmo_fluid}

We consider the simple case of a perfect fluid, where the equation of state is $w = P/\rho$, with $w$ a constant.  Solving the continuity equation \eqref{eq:continuity} gives us the energy density as a function of the scale factor:
\begin{equation}
	\rho(a) = \rho_0 a^{-3(1+w)},
\end{equation}
where $\rho(a_0) = \rho_0$, and $a_0 = 1$ as usual.

To investigate whether there is a big bounce, we insert this into Eq.~\eqref{eq:mod_friedmann_planck}, which becomes of the form
\begin{equation}
	\mathcal{H}^2 \propto a^{-3(1+w)} \left[
		1 - \frac{8 \pi \gamma^2 \tilde{\Delta}}{3}
		\frac{\rho_0}{\rho_{\rm Pl}} a^{\Theta} \right],
	\label{eq:mod_friedmann_fluid}
\end{equation}
where $\Theta$ depends on which regime of \eqref{eq:inv_vol_cases} we are in, namely
\begin{equation}
	\Theta = \left\{
\begin{aligned}
	& 2 \delta - 3(1+w),
		& {\rm if } \; a \gg a_{\star}
		    \\
	& 2 \delta + 4(2-l) - 3(1+w),
		& {\rm if } \; a \ll a_{\star}
\end{aligned}
    \right.
	    \label{eq:theta_bounce_cases}
\end{equation}
and we simply ignored the constant coefficients for $a \ll a_{\star}$. Whether a bounce happens depends on whether $\mathcal{H} \to 0$ when $a \neq 0$, which would happen if the higher-order correction in the modified Friedmann equation became dominant for small values of $a$, i.e. if $\Theta <0$.  The reason this is required is because $\rho$ needs to diverge faster than $\rho_c$ as $a \to 0$ in order for there to be a bounce.  This will happen when $w > w_b$, where
\begin{equation}
    w_b = \left\{
\begin{aligned}
	& -1 + \frac{2}{3} \delta,
		& {\rm if } \; a \gg a_{\star} \\
	& -1 + \frac{2}{3} \delta + \frac{4}{3}(2-l),
		& {\rm if } \; a \ll a_{\star}
\end{aligned}
    \right.
	    \label{eq:wb_bounce_cases}
\end{equation}
which means that, if the bounce does not happen in the $a \gg a_{\star}$ regime, the inverse-volume corrections make the bounce \emph{less} likely to happen.  If we use the favoured value of $\delta = 0$, and assume $l = 1$, then $w_b = 1/3$ and so $w$ still needs to be greater than that found for radiation in order for there to be a
bounce.  A possible candidate for this would be a massless (or kinetic-dominated) scalar field, where $w =1$.

Another aspect to investigate is whether the conditions for inflation are modified.  Taking \eqref{eq:eqn_state_acceleration}, we see that acceleration happens when $w<w_a$, where
\begin{equation}
	w_a = \left\{
\begin{aligned}
	& - \frac{1}{3} + \frac{16 \pi \gamma^2 \tilde{\Delta}}{9} (1-\delta) \frac{\rho_0}{\rho_{\rm Pl}} a^{\Theta},
    & {\rm if} \; a \gg a_{\star}
        \\
    & 1 - \frac{2l}{3} - \frac{16 \pi \gamma^2 \tilde{\Delta}}{a_{\star}^{4(2-l)}} \frac{1+\delta-l}{(1+l)^2} \frac{\rho_0}{\rho_{\rm Pl}} a^{\Theta},
    & {\rm if } \; a \ll a_{\star}
\end{aligned}
    \right.
	    \label{eq:wa_bounce_cases}
\end{equation}
so the range of values of $w$ which can cause accelerated expansion is indeed modified.  Holonomy-type corrections increase the range since we expect $\Theta < 0$, and so may inverse-volume corrections. However, the latter also seems to include a cut-off when the last term of Eq.~\eqref{eq:wa_bounce_cases} in the $a \ll a_{\star}$ regime dominates. Since a bounce requires $\dot{a}=0$ and $\ddot{a}>0$, the condition $w_b<w<w_a$ must be satisfied and so it must happen before the cut-off dominates if it is to happen at all.

%%%%%%%%%%%%%%%%%%%%%%%%%%%%%%%%%%%%%%%%%%%%%%%%%%%%%%%%%%%%%%%%%%%%%%

\subsection{Scalar field}
\label{subsec:cosmo_scalar}

We now investigate the effects that the inverse-volume and holonomy corrections can have when we have a scalar field.  In this case, the energy and pressure densities are given by
\begin{equation}
	\rho = \half \dot{\varphi}^2 + V(\varphi),
	\qquad
	P = \half \dot{\varphi}^2 - V(\varphi),
\end{equation}
and the continuity equation gives us the equation of motion for the scalar field,
\begin{equation}
	\ddot{\varphi} + 3 \mathcal{H} \dot{\varphi} + V' = 0,
	\label{eq:scalar_eom}
\end{equation}
where $V' := \partdif{V}{\varphi}$.

Let us investigate the era of slow-roll inflation.  Using the assumptions $|\ddot{\varphi}/V'| \ll 1$ and $\half \dot{\varphi}^2 \ll V$, we have the slow-roll equations,
\begin{subequations}
\begin{align}
	& \displaystyle{\dot{\varphi} = \frac{-V'}{3 \mathcal{H}}},
	\label{eq:slowroll_phidot}\\
	& \displaystyle{\mathcal{H}^2 = \frac{8 \pi
            \gamma^{\varnothing}}{3 m_{\rm Pl}^2} V \left( 1 - \frac{8
            \pi \gamma^2 \tilde{\Delta}}{3 m_{\rm Pl}^4} a^{2\delta}
          (\gamma^{\varnothing})^2 V \right)}.
	\label{eq:slowroll_hubble}
\end{align}
    \label{eq:slowroll}%
\end{subequations}
If we substitute \eqref{eq:slowroll_hubble} into \eqref{eq:slowroll_phidot}, take the derivative with respect to time and substitute in \eqref{eq:slowroll_hubble} and \eqref{eq:slowroll_phidot} again, we find
\begin{equation}
	\frac{\ddot{\varphi}}{V'} = \frac{1}{3} \eta,
	\qquad
	\frac{\dot{\varphi}^2}{2V} = \frac{1}{3} \epsilon,
\end{equation}
where the slow-roll parameters are
\begin{subequations}
\begin{align}
    \eta & :=
    \displaystyle{ \frac{1}{1-\sigma} \left( \frac{m_{\rm Pl}^2}{8\pi \gamma^{\varnothing}} \frac{V''}{V} -(1-2\sigma) \epsilon + \chi - \delta \sigma \right),}
        \label{eq:slowroll_eta} \\
    \epsilon & :=
    \displaystyle{ \frac{1}{1-\sigma} \frac{m_{\rm Pl}^2}{16\pi \gamma^{\varnothing}} \left(\frac{V'}{V}\right)^2, } 
        \label{eq:slowroll_epsilon} \\
    \chi & :=
    \displaystyle{ \frac{1-3\sigma}{2} \partdif{\,\ln \gamma^{\varnothing}}{\,\ln a} }
        \label{eq:slowroll_chi} \\
    \sigma & := 
    \displaystyle{ \frac{8 \pi \gamma^2 \tilde{\Delta}}{3m_{\rm Pl}^4} a^{2\delta} ( \gamma^{\varnothing} )^2 V,}
        \label{eq:slowroll_sigma}
\end{align}%
\end{subequations}
and the conditions for slow-roll inflation are
\begin{equation}
	|\eta | \ll 1,
        \quad
	\epsilon \ll 1,
        \quad
	|\chi | \ll 1,
        \quad
	|\sigma | \ll 1.
	\label{eq:slow_roll_parameters}
\end{equation}
We would like to investigate how these semi-classical effects affect the number of e-folds of the scale factor during inflation.  The number of e-folds before the end of inflation $\mathcal{N}(\varphi)$ is defined by $a(\varphi) = a_{\rm end} e^{-\mathcal{N}(\varphi)}$, where
\begin{equation}
	\mathcal{N}(\varphi)
	= -\int_{\varphi_{\rm end}}^{\varphi} {\rm d}\varphi
		\frac{\mathcal{H}}{\dot{\varphi}}
	= \frac{8\pi}{m_{\rm Pl}^2}
		\int_{\varphi_{\rm end}}^{\varphi} {\rm d}\varphi
		\frac{\gamma^{\varnothing} V}{V'}
		\left( 1-\frac{8 \pi \gamma^2 \tilde{\Delta}}{3m_{\rm Pl}^4}
		a^{2\delta}(\gamma^{\varnothing})^2 V \right).
\end{equation} 
If we remove the explicit dependence on $a$ from the integral by setting $\delta = 0$ and $\gamma^{\varnothing} = 1$ (i.e. taking only a certain form of holonomy corrections and ignoring inverse-volume corrections), and choose a power-law potential
\begin{equation}
	V(\varphi) = \frac{\lambda}{n} \varphi^n =
	\frac{\tilde{\lambda}}{n} m_{\rm Pl}^{4-n} \varphi^n,
	\label{eq:scalar_potential}
\end{equation}
where $\tilde{\lambda} >0$ and $n/2 \in \mathbb{N}$, then the number of e-folds before the end of inflation is
\begin{equation}
	\mathcal{N}(\varphi) = \frac{4 \pi}{n m_{\rm Pl}^2}
	\left( \varphi^2 - \varphi_{\rm end}^2 \right)
-\frac{64\pi^2\gamma^2\tilde{\Delta}\tilde{\lambda}}{3n^2(n+2)m_{\rm Pl}^{n+2}}
	\left( \varphi^{n+2} - \varphi_{\rm end}^{n+2} \right).
\end{equation}
If we take the approximation that slow-roll inflation is valid beyond the regime specified by \eqref{eq:slow_roll_parameters}, then we can calculate a value for the maximum amount of e-folds by starting inflation at the big bounce,
\begin{equation}
	\mathcal{N}_{\rm max} = \!\frac{4\pi}{n} \left[
	\left(\frac{3n}{8 \pi \gamma^2 \tilde{\Delta} \tilde{\lambda}} \right)^{2/n}
	\!-\left(\frac{\varphi_{\rm end}}{m_{\rm Pl}}\right)^2 \right]
	\!-\frac{64\pi^2\gamma^2\tilde{\Delta}\tilde{\lambda}}{3n^2(n+2)} \left[
	\left(\frac{3n}{8\pi\gamma^2\tilde{\Delta}\tilde{\lambda}}\right)^{1+2/n}
	\!- \left(\frac{\varphi_{\rm end}}{m_{\rm Pl}}\right)^{n+2} \right],
\end{equation}
and if we can assume $\varphi_{\rm end}/m_{\rm Pl} \ll 1$, then
\begin{equation}
	\mathcal{N}_{\rm max} = \frac{4\pi}{(n+2)} \left(
	\frac{3n}{8 \pi \gamma^2 \tilde{\Delta} \tilde{\lambda}} \right)^{2/n}.
\end{equation}

\begin{figure}[t]
	\begin{center}
	{\subfigure[Full phase space for $V(\varphi) = \lambda \varphi^2 /2$]{
		\label{fig:scalar_phase_2_full}
		\includegraphics[width = 0.333\textwidth]{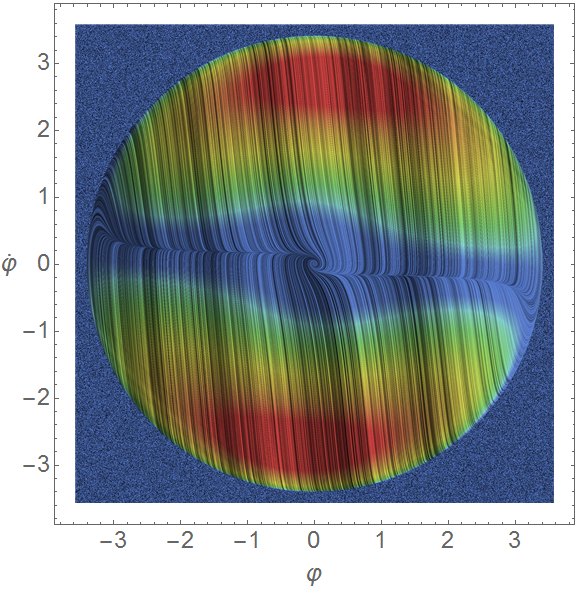}}}
	{\subfigure[Attractor solution for $V(\varphi) = \lambda \varphi^2 /2$]{
		\label{fig:scalar_phase_2_zoom}
		\includegraphics[width = 0.333\textwidth]{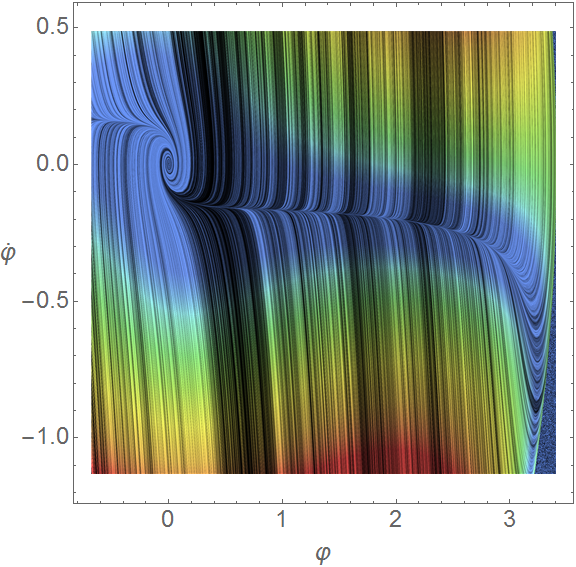}}}\\
	{\subfigure[Full phase space for $V(\varphi) = \lambda \varphi^4 /4$]{
		\label{fig:scalar_phase_4_full}
		\includegraphics[width = 0.333\textwidth]{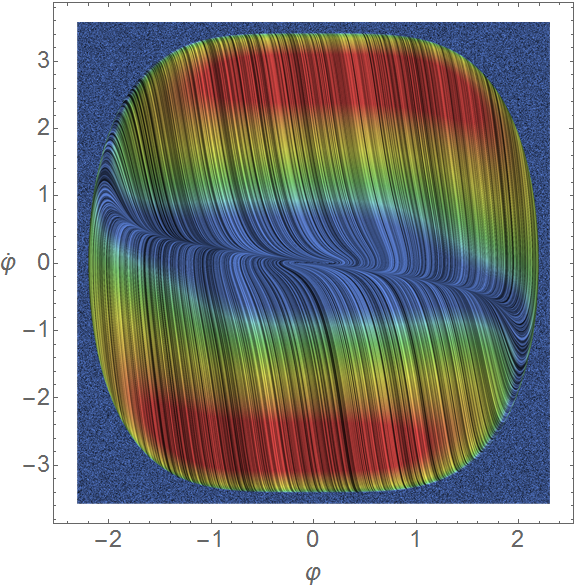}}}
	{\subfigure[Attractor solution for $V(\varphi) = \lambda \varphi^4 /4$]{
		\label{fig:scalar_phase_4_zoom}
		\includegraphics[width = 0.333\textwidth]{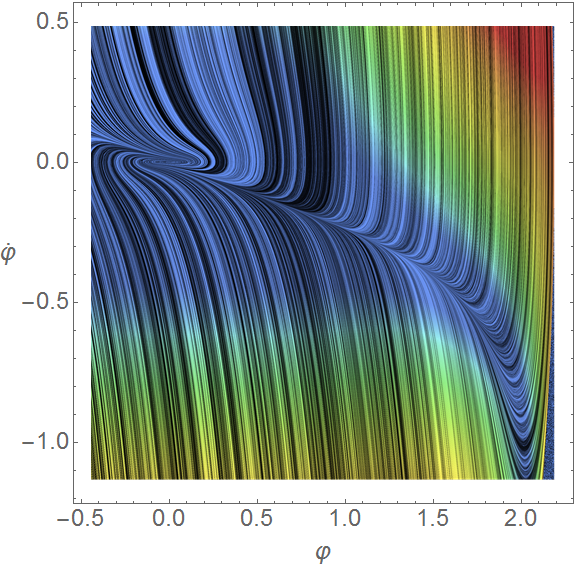}}}
	\end{center}
\caption{Line integral convolution plots showing trajectories in phase space for a scalar field with potential $\lambda \varphi^n / n$ with holonomy corrections.  The hue at each point indicates the magnitude of the vector $(\dot{\varphi},\ddot{\varphi})$, with blue indicating low values.  The trajectories do not extend outside of the region \eqref{eq:scalar_region}.  The attractor solution is well approximated by \eqref{eq:scalar_attractor}, corresponding to slow-roll inflation. We use $\tilde{\lambda} = 1$, $\tilde{\Delta} = 2 \sqrt{3} \pi \gamma$, $\delta = 0$, $\gamma^{\varnothing} = 1$, the plot is in Planck units.}
	\label{fig:scalar_phase}
\end{figure}

Let us now find the attractor solutions for slow-roll inflation. Substituting the Hubble parameter \eqref{eq:mod_friedmann_planck} into the equation of motion for the scalar field \eqref{eq:scalar_eom}, we
obtain
\begin{equation}
	\ddot{\varphi} + \dot{\varphi} \sqrt{ \frac{24 \pi
    \gamma^{\varnothing}}{m_{\rm Pl}^2} \left( \half \dot{\varphi}^2 +
  V \right) \left[ 1 - \frac{8 \pi \gamma^2 \tilde{\Delta}}{3 m_{\rm
        Pl}^4} a^{2 \delta} (\gamma^{\varnothing})^2 \left( \half
    \dot{\varphi}^2 + V \right) \right] } + V' = 0.
\end{equation}
We can remove the explicit scale-factor dependence of the equation by setting $\delta = 0$ and $\gamma^{\varnothing} = 1$ (the same assumptions as we used to find $\mathcal{N}$).  Then substituting in the
power-law potential \eqref{eq:scalar_potential} we get
\begin{equation}
	\ddot{\varphi} + \dot{\varphi} \sqrt{ \frac{24 \pi}{m_{\rm Pl}^2}
  \left( \half \dot{\varphi}^2 + \frac{\lambda}{n} \varphi^n \right)
  \left[ 1 - \frac{8 \pi \gamma^2 \tilde{\Delta}}{3 m_{\rm Pl}^4}
    \left( \half \dot{\varphi}^2 + \frac{\lambda}{n} \varphi^n \right)
    \right] } + \lambda \varphi^{n-1} = 0,
\end{equation}
which is applicable only for the region $\rho < \rho_c$, or 
\begin{equation}
	1 - \frac{8 \pi \gamma^2 \tilde{\Delta}}{3 m_{\rm Pl}^4} \left( \half
\dot{\varphi}^2 + \frac{\lambda}{n} \varphi^n \right) >0,
	\label{eq:scalar_region}
\end{equation}
otherwise $\mathcal{H}$ is complex.  We use this equation to plot phase space trajectories in \figref{fig:scalar_phase}.

We can find the slow-roll attractor solution for $|\ddot{\varphi} \varphi^{1-n} / \lambda| \ll 1$ and $\half \dot{\varphi}^2 \ll \frac{\lambda}{n} \varphi^n$,
\begin{equation}
	\dot{\varphi} \approx - \sqrt{\frac{n \lambda m_{\rm Pl}^2}{24
    \pi}} \varphi^{\frac{n}{2}-1} \left( 1 - \frac{8 \pi \gamma^2
  \tilde{\Delta} \lambda}{3 n m_{\rm Pl}^4} \varphi^n \right)^{-1/2},
	\label{eq:scalar_attractor}
\end{equation}
where the term in the bracket is the correction to the classical solution.  Looking at Figs.~\ref{fig:scalar_phase_2_zoom} and \ref{fig:scalar_phase_4_zoom}, we conclude that the attractor
solutions diverge from a linear relationship as they approach the boundary.

\begin{figure}[t]
	\begin{center}
	{\subfigure[Accelerating values of $w$ for $V(\varphi) =
            \lambda \varphi^2 /2$]{
		\label{fig:scalar_eos_2}
		\includegraphics[width = 0.333\textwidth]{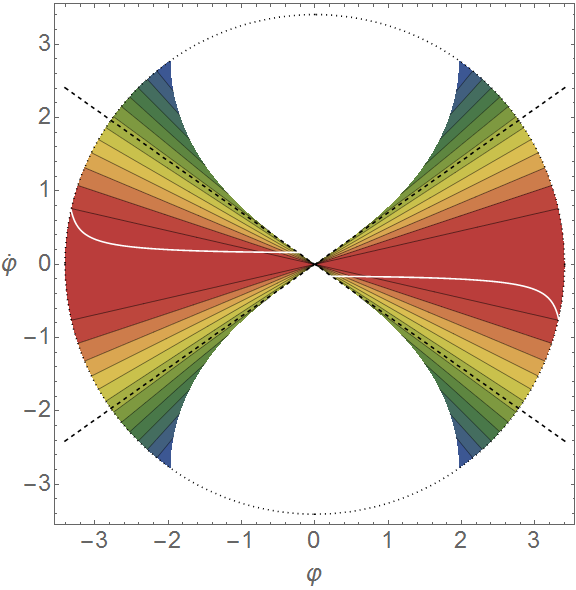}}}
        {\subfigure[Accelerating values of $w$ for $V(\varphi) =
            \lambda \varphi^4 /4$]{
		\label{fig:scalar_eos4}
		\includegraphics[width = 0.333\textwidth]{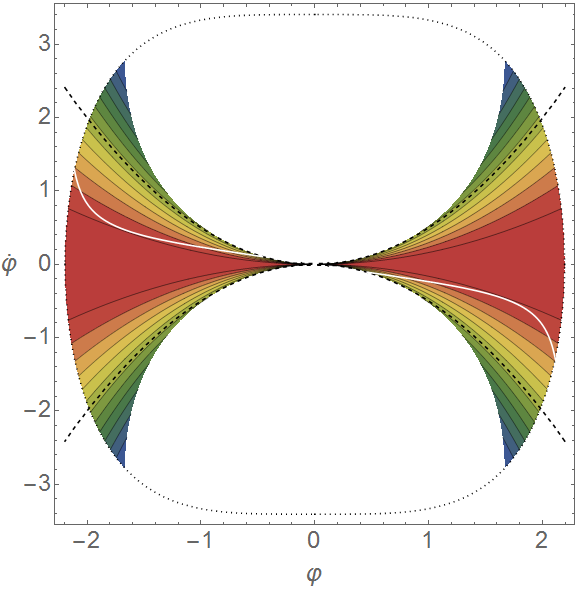}}}
	\end{center}
\caption{Contour plots showing the region in scalar phase space satisfying the condition for accelerated expansion when holonomy corrections are included \eqref{eq:scalar_acceleration}.  The dashed line indicates the classical acceleration condition $w_a = -1/3$ and the dotted line indicates the bounce boundary.  The white line indicates the slow-roll solution \eqref{eq:scalar_attractor}. The contours indicate the value of $w$ by their colour, and the most blue contour is for $w \approx 0.2$. We use $\tilde{\lambda} = 1$, $\tilde{\Delta} = 2 \sqrt{3} \pi \gamma$, $\delta = 0$, $\gamma^{\varnothing} = 1$, the plot is in Planck units.}
	\label{fig:scalar_eos}
\end{figure}

The condition for acceleration for the case we are considering here is
\begin{equation}
	w < w_{a} = \frac{-1}{3} \left[
		1 - \frac{16 \pi \gamma^2 \tilde{\Delta}}{3 m_{\rm Pl}^4} \left(
			\half \dot{\varphi}^2 + \frac{\lambda}{n} \varphi^n
		\right)
	\right]
	\label{eq:scalar_acceleration}
\end{equation}
we plot in \figref{fig:scalar_eos} this region on the phase space of the scalar field to see how accelerated expansion can happen in a wider range than in the classical case.  In order to be able to solve the equations and make plots, we have neglected non-zero values of $\delta$ and non-unity values of $\gamma^{\varnothing}$. It may be that in these cases the big bounce and inflation are no longer inevitable, as it was found for the perfect fluid.

%%%%%%%%%%%%%%%%%%%%%%%%%%%%%%%%%%%%%%%%%%%%%%%%%%%%%%%%%%%%%%%%%%%

\subsection{Cosmology discussion}
\label{subsec:cosmo_discussion}

We found that higher curvature corrections (that are likely to arise due to holonomy corrections) are those responsible for the repulsive gravitational effect which produce the big bounce.  For a perfect fluid, the effects that the quantum corrections have depend on the equation of state, but inflation and a big bounce are possible.

For a scalar field, we found the slow-roll conditions and the equation for the number of e-folds of inflation.  However, to find a simple solution we restricted ourselves to a certain form of holonomy corrections.  By plotting the region of phase space which allows acceleration, we showed that holonomy corrections aid inflation and
they make the big bounce inevitable.  How the results are affected when inverse-volume corrections are included has not been discussed.

%%%%%%%%%%%%%%%%%%%%%%%%%%%%%%%%%%%%%%%%%%%%%%%%%%%%%%%%%%%%%%%%

\section{Conclusions}
\label{sec:concl}

We found that we can regain an effective Lagrangian to fourth order in extrinsic curvature from a general deformation of the algebra of constraints.  The flexibility gained from having an effective theory
which is not already symmetry-reduced may be valuable for corroborating results across different types of systems.  We have applied our results to isotropic early universe cosmology, and looked for bounce and inflation conditions for certain matter contents. However, it could in principle also be used to study spherically
symmetric models or anisotropic cosmology. Let us again emphasise that in our analysis we only keep holonomy corrections manifesting as higher-order powers of curvature and ignore higher-order terms in the derivative expansion.

%One needs to bear in mind that the higher-order curvature terms we have been dealing with are just of the extrinsic type, and so general covariance is broken.  Results are observer dependent.
One needs to bear in mind that the higher-order curvature terms we have been dealing with are just of the extrinsic type, and so general covariance is deformed. Results remain observer independent, but one would have to correct the classical equations for transformations between different frames.
However, as stated previously, it may be possible to change variables to absorb quantum corrections and regain undeformed general covariance \cite{tibrewala_inhomogeneities_2013}.  We may not be able to perform this transformation with our Lagrangian because canonical transformations require a Hamiltonian defined in terms of the momentum variable.  In Ref.~\cite{hojman_geometrodynamics_1976}, a similar calculation was performed to regain the gravitational Hamiltonian rather than the Lagrangian, so an extension of this calculation which included the deformations might then allow us to explore this transformation of variables.

The effective Lagrangian we got from the deformed constraint algebra does not specify where the corrections come from.  We simply worked in a canonical gravity scheme with a deformed symmetry using geometrodynamical variables. Since there are other approaches to quantum gravity which are also rooted in canonical methods, it is possible that our results are more general than only being relevant to loop quantum gravity.  We may look into how we could use our results to compare different theories.

%%%%%%%%%%%%%%%%%%%%%%%%%%%%%%%%%%%%%%%%%%%%%%%%%%%%%%%%%%%%%%%%%

\appendix

\section{Non-locality}
\label{sec:nonlocal}

If the underlying quantum gravity theory is discrete, there will necessarily be some non-local effects when we try to approximate it using a classical and continuous manifold.  In loop quantum gravity, these come from the quantised holonomies, since they are classically equivalent to path-ordered exponentials of the connection variable integrated along an unshrinkable path.  In order to find the semi-classical effects from this apparent non-locality, we re-sum over all derivatives of $v_{ab}$ in the expansions \eqref{eq:local_expansion}, to get
\begin{subequations}
\begin{align}
	L(x) & = L^{\varnothing}(x) + \sum^{\infty}_{m=0} \left[ \left( \sum^{\infty}_{n_0=0} \cdots \sum^{\infty}_{n_m=0} L_{(n_0, \ldots, n_m)}(x) \right) - L^{\varnothing}(x) \right],
		\label{eq:nonlocal_lagrangian_expansion}
	    \\
	\beta(x) & = \beta^{\varnothing}(x) + \sum^{\infty}_{m=0} \left[ \left( \sum^{\infty}_{n_0=0} \cdots \sum^{\infty}_{n_m=0} \beta_{(n_0, \ldots, n_m)}(x) \right) - \beta^{\varnothing}(x) \right],
		\label{eq:nonlocal_correction_expansion}
\end{align}
	\label{eq:nonlocal_expansion}%
\end{subequations}
where each $n_i$ in $L_{(n_0,\ldots,n_m)}$ and $\beta_{(n_0,\ldots,n_m)}$ is the number of $i$th derivatives of
$v_{ab}$ which these terms contain.  For example, 
\begin{equation}
    L_{(1,1)}[q_{ij}, v_{ij}] = L^{A B b_3}[q_{ij}] v_{A} v_{B|b_3}.
\end{equation}
We have included the extra $L^{\varnothing}$ and $\beta^{\varnothing}$ in \eqref{eq:nonlocal_expansion} in order to not count these terms multiple times, since $L^{\varnothing} := L_{(0)} = L_{(0,0)} = \dots$

If we find the Lagrangian expansion to second order of derivatives (remembering that each factor of $v_{ab}$ implicitly contains a time derivative), we get
\begin{equation}
	L=L^{\varnothing} + L^A v_A + L^{A a_3} v_{A|a_3}
	+ L^{AB} v_{A} v_{B}.
	\label{eq:nonlocal_lagrangian_expansion_2nd}
\end{equation}
Our distribution equation \eqref{eq:dist_eqn} is not adequate here, as the partial derivatives with respect to $v_{ab}$ must be replaced with functional derivatives in order to include the dependence on derivatives of $v_{ab}$.  Our distribution equation is now,
\begin{equation}
	\funcdif{L(x)}{q_{A}(y)} v_{A}(y) \delta(x,z)
	+ 2 \beta_{|a_2}(x) \funcdif{L(x)}{v_{A}(z)} \delta_{|a_1}(x,y)
	+ 2 \beta(x) \funcdif{L(x)}{v_{A}(z)} \delta_{|A}(x,y)
	- (x \leftrightarrow y) = 0.
	\label{eq:nl_dist_eqn}
\end{equation}
Substituting in the non-local expansion \eqref{eq:nonlocal_lagrangian_expansion_2nd} and setting $v_{ab}=0$, we get
\begin{equation}
	2 \left\{
		L^A(x) \delta(x,z) + L^{A a_3}(x) \delta_{|a_3}(x,z)
	\right\} \left[
		\beta^{\varnothing}(x) \delta_{|a_1}(x,y)	
	\right]_{|a_2}^{(x)} - (x \leftrightarrow y) = 0.
\end{equation}
Integrating this equation over $z$, the term proportional to $L^{Aa_3}$ becomes a total derivative and we can thus discard it.  The remaining equation simply leads us to \eqref{eq:lagrangian_identities_1}.

Going back to Eq.~\eqref{eq:nl_dist_eqn}, substituting in Eq.~\eqref{eq:nonlocal_lagrangian_expansion_2nd}, taking functional derivatives with respect to $v_C(w)$ and then setting $v_A = 0$, we get
\begin{equation}
\begin{split}
	0 & = \funcdif{L^{\varnothing}(x)}{q_C(y)} \delta(y,w)
        \\
	& + 2 \delta_{|a_1}(x,y) \left\{ L^A(x) \left[ \beta^C(x) \delta(x,w) + \beta^{C c_3}(x) \delta_{|c_3}(x,w) \right]_{|a_2} + 2 \beta^{\varnothing}_{|a_2}(x) L^{AC}(x) \delta(x,w) \right\}
        \\
	& + 2 \delta_{|A}(x,y) \left\{ L^A(x) \left[ \beta^C(x) \delta(x,w) + \beta^{C c_3}(x) \delta_{|c_3}(x,w) \right] + 2 \beta^{\varnothing}(x) L^{AC}(x) \delta(x,w) \right\}
        \\
	& - (x \leftrightarrow y).
\end{split}
\end{equation}
Now, we move the derivatives by using the product rule so that $\delta(x,w)$ is not differentiated and then discard total derivative terms.  If we use \eqref{eq:lagrangian_identities_1}, use the product rule to distribute derivatives and see what terms cancel, we are left with
\begin{equation}
	0 = \left\{
		-\funcdif{L^{\varnothing}(x)}{q_C(y)}
		+4\beta^{\varnothing}_{|a_2}(x) L^{AC}(x) \delta_{|a_1}(x,y)
		+4\beta^{\varnothing}(x) L^{AC}(x) \delta_{|A}(x,y)
	\right\} \delta(x,w) - (x \leftrightarrow y),
\end{equation}
which is the same as equation $(60)$ in Ref.~\cite{bojowald_deformed_2012}. Therefore, it will lead to the same effective Lagrangian at second order as the local case \eqref{eq:lagrangian_2nd}.  Therefore, non-local effects do not appear at second order.

To third order in derivatives, the Lagrangian looks like
\begin{equation}
	L=L^{\varnothing} + L^A v_A + L^{A a_3} v_{A|a_3}
	+ L^{A a_3 a_4} v_{A|a_3 a_4}
	+ L^{AB} v_{A} v_{B} + L^{AB b_3} v_{A} v_{B|b_3}
	+ L^{ABC} v_{A} v_{B} v_{C},
\end{equation}
so there are two more terms to consider compared to the second order case.  The calculation of how this may affect the effective Lagrangian is not considered here.

%%%%%%%%%%%%%%%%%%%%%%%%%%%%%%%%%%%%%%%%%%%%%%%%%%%%%%%%%%%%%%%%%

\section{Extra 3rd-order calculations}
\label{sec:3rd_extra}

These calculations are not crucial to our 3rd-order Lagrangian calculations, but they do help give justification for why we think that the first order term $L^A$ vanishes.

Let us start from Eq.~\eqref{eq:dist_eqn_3rd} and follow the method of Ref.~\cite{kuchar_geometrodynamics_1974}. We decompose $v_{ab}$ (which can be treated as an arbitrary function) into scalar and tensor components $v_{ab}(x) = \bar{v}(x) \nu_{ab}(x)$, which we can vary independently (making sure that we keep $\det \nu = 1$).  This means that Eq.~\eqref{eq:dist_eqn_3rd} can be rewritten as
\begin{equation}
    0 = A(x,y) \bar{v}^2 (x) - A(y,x) \bar{v}^2 (y) + B(x,y) \bar{v}(x) \bar{v}(y) + C^{a}(x,y) \bar{v}(x) \bar{v}_{|a}(x) - C^{a}(y,x) \bar{v}(y) \bar{v}_{|a}(y),
    	\label{eq:dist_eqn_3rd_split}
\end{equation}
where
\begin{subequations}
\begin{align}
    & A (x,y) = A^{a_1}_1 (x) \delta_{|a_1} (x,y) + A^{A}_2 (x) \delta_{|A} (x,y),
        \\
    & A^{a_1}_1 = 2 \left( L^{A} \beta^{BC}_{|a_2} + 2 L^{AB} \beta^{C}_{|a_2} \right) \nu_{B} \nu_{C} + 2 \left( L^{A} \beta^{BC} + L^{AB} \beta^{C} \right) (\nu_{B} \nu_{C})_{|a_2}, 
        \\
    & A^{A}_2 = 2 \left( L^{A} \beta^{BC} + 2 L^{AB} \beta^{C} + 3 L^{ABC} \beta^{\varnothing} \right) \nu_{B} \nu_{C},
\end{align}
	\label{eq:dist_eqn_3rd_split_Adef}%
\end{subequations}
\begin{gather}
    B(x,y) = \left( \funcdif{L^{A}(x)}{q_{B}(y)} - \funcdif{L^{B}(y)}{q_{A}(x)} \right) \nu_{A}(x) \nu_{B}(y),
	    \label{eq:dist_eqn_3rd_split_Bdef}
	        \\
    C^{a_1}(x,y) = C^{A}_1(x) \delta_{|a_2}(x,y),
        \qquad
    C^{A}_1 = 4 \left( L^{A} \beta^{BC} + L^{AB} \beta^{C} \right) \nu_{B} \nu_{C}.
	    \label{eq:dist_eqn_3rd_split_Cdef}
\end{gather}
Setting $\bar{v}_{|a} = 0$ in Eq.~\eqref{eq:dist_eqn_3rd_split}, we
find 
\begin{equation}
    0 = A(x,y) \bar{v}^2 (x) - A(y,x) \bar{v}^2 (y) + B(x,y) \bar{v}(x) \bar{v}(y),
	    \label{eq:dist_eqn_3rd_split_AB}
\end{equation}
which means that
\begin{equation}
    0 = C^{a}(x,y) \bar{v}(x) \bar{v}_{|a}(x) - C^{a}(y,x) \bar{v}(y) \bar{v}_{|a}(y),
	    \label{eq:dist_eqn_3rd_split_C}
\end{equation}
must be satisfied independently.  

We can easily show that \eqref{eq:dist_eqn_3rd_split_AB} is satisfied by following a similar procedure to Ref.~\cite{kuchar_geometrodynamics_1974}, but it does not give any useful new conditions on the expansion coefficients of $L$ or $\beta$.  So we turn our attention to \eqref{eq:dist_eqn_3rd_split_C}.  We take functional derivatives with respect to $\bar{v}(z)$ and $\bar{v}(z')$, multiply by test functions $a(y)$, $b(z)$, $c(z')$ and integrate by parts over $z'$,
\begin{equation}
\begin{split}
    0 & = a(y) b(z) C^{ab}_{1}(x) \delta_{|b}(x,y) \left[ \delta_{|a}(x,z) c(x) - \delta(x,z) c_{|a}(x) \right]
        \\
    & - a(y) b(z) C^{ab}_{1}(y) \delta_{|b}(x,y) \left[ \delta_{|a}(y,z) c(y) - \delta(y,z) c_{|a}(y) \right],
\end{split}
\end{equation}
then integrate by parts over $y$ and discard terms which vanish due to symmetry of indices,
\begin{equation}
\begin{split}
    0 & = a(x) b(z) \left\{ -\delta(x,z) \left[ C^{ab}_{1|b}(x) c_{|a}(x) + C^{ab}_1 (x) c_{|ab}(x) \right] \right.
        \\
    & \left. + \delta_{|a} (x,z) C^{ab}_{1|b}(x) c(x) + \delta_{|ab}(x,z) C^{ab}_1(x) c(x) \right\},
\end{split}
\end{equation}
then, integrate by parts over $z$, set $a(x) = 1$, and then integrate by parts over $x$ to get
\begin{equation}
    0 = - 2 \int {\rm d}^3 x b_{|a}(x) c(x) C^{ab}_{1|b}(x),
\end{equation}
and since $b(x)$ and $c(x)$ are arbitrary test functions, we find $C^{ab}_{1|b} = 0$, and thus
\begin{equation}
    0 = \left( L^A \beta^{BC} + L^{AB} \beta^{C} \right)_{|a_2} \nu_{B} \nu_{C} + \left( L^A \beta^{BC} + L^{AB} \beta^{C} \right) \left( \nu_B \nu_C \right)_{|a_2},
\end{equation}
and by remembering that $\nu_{A}$ is an arbitrary function, we see that both terms must vanish independently, and so we find 
\begin{equation}
	L^A \beta^{BC} + L^{AB} \beta^C = 0.
	    \label{eq:lagrangian_identities_3extra}
\end{equation}
This is necessary for our argument in Section~\ref{subsubsec:regainlag_3rd_consistency}.

%%%%%%%%%%%%%%%%%%%%%%%%%%%%%%%%%%%%%%%%%%%%%%%%%%%%%%%%%%%%%%%%%

\begin{acknowledgments}
It is a pleasure to thank Martin Bojowald for helpful discussions about this work.  The work of RC is supported by an STFC studentship.
\end{acknowledgments}

%%%%%%%%%%%%%%%%%%%%%%%%%%%%%%%%%%%%%%%%%%%%%%%%%%%%%%%%%%%%%%%%%

\bibliographystyle{./4odgrbib}
\bibliography{./4odgr}

\providecommand{\href}[2]{#2}\begingroup\raggedright\begin{thebibliography}{10}

\bibitem{bojowald2010canonical}
M.~Bojowald, {\em {Canonical Gravity and Applications: Cosmology, Black Holes,
  and Quantum Gravity}}.
\newblock Cambridge University Press, 2010.

\bibitem{Rovelli2014}
C.~Rovelli and F.~Vidotto, {\em {Introduction to covariant loop quantum
  gravity}}.
\newblock Cambridge University Press, to appear, 2015.

\bibitem{teitelboim_how_1973}
C.~Teitelboim, ``{How Commutators of Constraints Reflect Spacetime
  Structure},'' {\em Ann. Phys. (N.Y.)} {\bf 79} (1973) 542--557.

\bibitem{bojowald_loop_2004}
M.~Bojowald, ``{Loop Quantum Cosmology: Recent Progress},'' {\em Pramana} {\bf
  63} (2004) 765--776, [\href{http://xxx.lanl.gov/abs/gr-qc/0402053}{{\tt
  gr-qc/0402053}}].

\bibitem{Ashtekar2006}
A.~Ashtekar, T.~Pawlowski, and P.~Singh, ``{Quantum nature of the big bang:
  Improved dynamics},'' {\em Phys. Rev.} {\bf D74} (2006) 084003,
  [\href{http://xxx.lanl.gov/abs/gr-qc/0607039}{{\tt gr-qc/0607039}}].

\bibitem{sakellariadou_lattice_2007-1}
W.~Nelson and M.~Sakellariadou, ``{Lattice Refining Loop Quantum Cosmology and
  Inflation},'' {\em Phys. Rev.} {\bf D76} (2007) 044015,
  [\href{http://xxx.lanl.gov/abs/0706.0179}{{\tt arXiv:0706.0179}}].

\bibitem{bojowald_quantum_2012}
M.~Bojowald, ``{Quantum Cosmology: Effective Theory},'' {\em Class. Quant.
  Grav.} {\bf 29} (2012) 213001, [\href{http://xxx.lanl.gov/abs/1209.3403}{{\tt
  arXiv:1209.3403}}].

\bibitem{Cailleteau2012a}
T.~Cailleteau, J.~Mielczarek, A.~Barrau, and J.~Grain, ``{Anomaly-free scalar
  perturbations with holonomy corrections in loop quantum cosmology},'' {\em
  Class. Quant. Grav.} {\bf 29} (2012) 095010,
  [\href{http://xxx.lanl.gov/abs/1111.3535}{{\tt arXiv:1111.3535}}].

\bibitem{Cailleteau2013}
T.~Cailleteau, L.~Linsefors, and A.~Barrau, ``{Anomaly-free perturbations with
  inverse-volume and holonomy corrections in Loop Quantum Cosmology},'' {\em
  Class. Quant. Grav.} {\bf 31} (2014) 125011,
  [\href{http://xxx.lanl.gov/abs/1307.5238}{{\tt arXiv:1307.5238}}].

\bibitem{bojowald_deformed_2012}
M.~Bojowald and G.~M. Paily, ``{Deformed General Relativity and Effective
  Actions from Loop Quantum Gravity},'' {\em Phys. Rev.} {\bf D86} (2012)
  104018, [\href{http://xxx.lanl.gov/abs/1112.1899}{{\tt arXiv:1112.1899}}].

\bibitem{magueijo_lorentz_2002}
J.~Magueijo and L.~Smolin, ``{Lorentz invariance with an invariant energy
  scale},'' {\em Phys. Rev. Lett.} {\bf 88} (2002) 190403,
  [\href{http://xxx.lanl.gov/abs/hep-th/0112090}{{\tt hep-th/0112090}}].

\bibitem{magueijo_generalized_2003}
J.~Magueijo and L.~Smolin, ``{Generalized Lorentz invariance with an invariant
  energy scale},'' {\em Phys. Rev.} {\bf D67} (2003) 044017,
  [\href{http://xxx.lanl.gov/abs/gr-qc/0207085}{{\tt gr-qc/0207085}}].

\bibitem{bojowald_loop_2008}
M.~Bojowald and G.~M. Hossain, ``{Loop quantum gravity corrections to
  gravitational wave dispersion},'' {\em Phys. Rev.} {\bf D77} (2008) 023508,
  [\href{http://xxx.lanl.gov/abs/0709.2365}{{\tt arXiv:0709.2365}}].

\bibitem{hossenfelder_box-problem_2009}
S.~Hossenfelder, ``{The Box-Problem in Deformed Special Relativity},''
  \href{http://xxx.lanl.gov/abs/0912.0090}{{\tt arXiv:0912.0090}}.

\bibitem{hossenfelder_multi-particle_2007}
S.~Hossenfelder, ``{Multi-Particle States in Deformed Special Relativity},''
  {\em Phys. Rev.} {\bf D75} (2007) 105005,
  [\href{http://xxx.lanl.gov/abs/hep-th/0702016}{{\tt hep-th/0702016}}].

\bibitem{mielczarek_asymptotic_2012}
J.~Mielczarek, ``{Asymptotic silence in loop quantum cosmology},'' {\em AIP
  Conf. Proc.} {\bf 1514} (2012) 81--84,
  [\href{http://xxx.lanl.gov/abs/1212.3527}{{\tt arXiv:1212.3527}}].

\bibitem{ambjorn_quantum_2013}
J.~Ambjorn, A.~Goerlich, J.~Jurkiewicz, and R.~Loll, ``{Quantum Gravity via
  Causal Dynamical Triangulations},''
  \href{http://xxx.lanl.gov/abs/1302.2173}{{\tt arXiv:1302.2173}}.

\bibitem{Hartle1983}
J.~Hartle and S.~Hawking, ``{Wave function of the Universe},'' {\em Phys. Rev.}
  {\bf D28} (1983) 2960--2975.

\bibitem{tibrewala_inhomogeneities_2013}
R.~Tibrewala, ``{Inhomogeneities, loop quantum gravity corrections, constraint
  algebra and general covariance},'' {\em Class. Quant. Grav.} {\bf 31} (2014)
  055010, [\href{http://xxx.lanl.gov/abs/1311.1297}{{\tt arXiv:1311.1297}}].

\bibitem{hojman_geometrodynamics_1976}
S.~Hojman, K.~Kuchar, and C.~Teitelboim, ``{Geometrodynamics Regained},'' {\em
  Ann. Phys. (N.Y.)} {\bf 96} (1976) 88--135.

\bibitem{kuchar_geometrodynamics_1974}
K.~V. Kuchar, ``{Geometrodynamics regained: a Lagrangian approach},'' {\em J.
  Math. Phys.} {\bf 15} (1974) 708--715.

\bibitem{Deruelle2010}
N.~Deruelle, M.~Sasaki, Y.~Sendouda, and D.~Yamauchi, ``{Hamiltonian
  Formulation of f (Riemann) Theories of Gravity},'' {\em Prog. Theor. Phys.}
  {\bf 123} (2010) 169--185, [\href{http://xxx.lanl.gov/abs/0908.0679}{{\tt
  arXiv:0908.0679}}].

\bibitem{bojowald_discreteness_2014}
M.~Bojowald, G.~M. Paily, and J.~D. Reyes, ``{Discreteness corrections and
  higher spatial derivatives in effective canonical quantum gravity},'' {\em
  Phys. Rev.} {\bf D90} (2014) 025025,
  [\href{http://xxx.lanl.gov/abs/1402.5130}{{\tt arXiv:1402.5130}}].

\bibitem{bojowald_radiation_2007}
M.~Bojowald and R.~Das, ``{The radiation equation of state and loop quantum
  gravity corrections},'' {\em Phys. Rev.} {\bf D75} (2007)
  [\href{http://xxx.lanl.gov/abs/0710.5721}{{\tt arXiv:0710.5721}}].

\bibitem{bojowald_quantum_2013}
M.~Bojowald, G.~M. Hossain, M.~Kagan, and C.~Tomlin, ``{Quantum matter in
  quantum space-time},'' {\em Quantum Matter} {\bf 2} (2013) 436--443,
  [\href{http://xxx.lanl.gov/abs/1302.5695}{{\tt arXiv:1302.5695}}].

\bibitem{sakellariadou_lattice_2007}
W.~Nelson and M.~Sakellariadou, ``{Lattice Refining {LQC} and the Matter
  Hamiltonian},'' {\em Phys. Rev.} {\bf D76} (2007) 104003,
  [\href{http://xxx.lanl.gov/abs/0707.0588}{{\tt arXiv:0707.0588}}].

\end{thebibliography}\endgroup

\end{document}